\documentclass{article}
\usepackage[margin=2cm]{geometry}

\usepackage{float}

\usepackage{changepage}

\usepackage[utf8]{inputenc}

\usepackage{textcomp,marvosym}

\usepackage{fixltx2e}

\usepackage{amsmath,amssymb}

\usepackage{cite}

\usepackage{nameref,hyperref}

\usepackage{microtype}
\DisableLigatures[f]{encoding = *, family = * }

\usepackage{rotating}

\usepackage[aboveskip=1pt,labelfont=bf,labelsep=period,justification=raggedright,singlelinecheck=off]{caption}


\DeclareMathOperator{\E}{\mathbb{E}}
\DeclareMathOperator{\Proba}{\mathbb{P}}

\newcommand{\Eb}[1]{\ensuremath{\E\!\left[#1\right]}}
\newcommand{\Pb}[1]{\ensuremath{\Proba\!\left[#1\right]}}

\usepackage{ragged2e}

\title{Calibration of a Density-based Model of Urban Morphogenesis}
\date{}

\author{Raimbault Juste\textsuperscript{1,2,*}\\
\textsuperscript{1}UMR CNRS 8504 G{\'e}ographie-cit{\'e}s, Paris, France\\
\textsuperscript{2}UMR-T 9403 IFSTTAR LVMT, Champs-sur-Marne, France\\
Email : juste.raimbault@polytechnique.edu}

\begin{document}

\maketitle

\begin{abstract}
\justify
We study a stochastic model of urban growth generating spatial distributions of population densities at an intermediate mesoscopic scale. The model is based on the antagonist interplay between the two opposite abstract processes of aggregation (preferential attachment) and diffusion (urban sprawl). Introducing indicators to quantify precisely urban form, the model is first statistically validated and intensively explored to understand its complex behavior across the parameter space. We then compute real morphological measures on local areas of size 50km covering all European Union, and show that the model can reproduce most of existing urban morphologies in Europe. It implies that the morphological dimension of urban growth processes at this scale are sufficiently captured by the two abstract processes of aggregation and diffusion.
\end{abstract}

\justify

%%%%%%%%%%%%%%%%%%%%%%%%
\section*{Introduction}

The study of urban growth, and more particularly its quantification, is more than ever a crucial issue in a context where most of the world population live in cities which expansion has significant environmental impacts~\cite{seto2012global} and that have therefore to ensure an increased sustainability and resilience to climate change. The understanding of drivers for urban growth can lead to better integrated policies. It is however a question far from being solved in the diverse related disciplines: Urban Systems are complex socio-technical systems that can be studied from a large variety of viewpoints. Batty has advocated in that sense for the construction of a dedicated science defined by its objects of study more than the methods used~\cite{batty2013new}, what would allow easier coupling of approaches and therefore Urban Growth models taking into account heterogeneous processes. The processes that a model can grasp are also linked to the choice of the scale of study. At a macroscopic scale, models of growth in system of cities are mainly the concern of economics and geography. \cite{gabaix1999zipf} shows that in first approximation, the Gibrat's model postulating random growth rates not depending on city size, yield the well-know Zipf's law, or rank-size law, which is a typical stylized fact witnessing hierarchy in systems of cities. It was however shown empirically that systematic deviations to this law exist~\cite{rozenfeld2008laws}, and that spatial interactions may be responsible for it. Models integrating spatial interactions include for example \cite{bretagnolle2000long} that introduces a growth model in which these interactions, that are function of distance and the geography, play a significant role in growth rates. More recently, \cite{favaro2011gibrat} has extended this model by taking into account innovation waves between cities as a driver. The interplay of space, economic and population growth is studied by the Marius model~\cite{cottineau2014evolution} in the case of the former Soviet Union, on which model performance is shown improved compared to models without interactions.

At smaller scales, that can be understood as microscopic or mesoscopic depending on the resolution and extent of models, agents of models fundamentally differ. Space is generally taken into account in a finer way, through neighborhood effects for example. For example, \cite{andersson2002urban} propose a micro-based model of urban growth, with the purpose to replace non-interpretable physical mechanisms with agent mechanisms, including interactions forces and mobility choices. Local correlations are used in \cite{makse1998modeling}, which develops the model introduced in~\cite{makse1995modelling}, to modulate growth patterns to ressemble real configurations. The world of Cellular Automata (CA) models of Urban Growth~\cite{batty1994cells} also offers numerous examples. \cite{GEAN:GEAN940} introduced a generic framework for CA with multiple land use, based on local evolution rules. A model with simpler states (occupied or not) but taking into account global constraints is studied by  \cite{ward2000stochastically}. The Sleuth model, initially introduced by \cite{clarke1998loose} for the San Francisco Bay area, and for which an overview of diverse applications is given in~\cite{clarke2007decade}, was calibrated on areas all over the world, yielding comparative measures through the calibrated parameters.

Closely related to CA models but not exactly similar are Urban Morphogenesis models, which aim to simulate the growth of urban form from autonomous rules. \cite{frankhauser1998fractal} suggested that the fractal nature of cities is closely to the emergence of the form from the microscopic socio-economic interactions, namely urban morphogenesis. \cite{courtat2011mathematics} develops a morphogenesis model for urban roads alone, with growth rules based on geometrical considerations. These are shown sufficient to produce a broad range of patterns analog to existing ones. Similarly, \cite{raimbault2014hybrid} couples a CA with an evolving network to reproduce stylized urban form, from concentrated monocentric cities to sprawled suburbs. The Diffusion-Limited-Aggregation model, coming from physics, and which was first studied for cities by~\cite{batty1991generating}, can also be seen as a morphogenesis model. These kind of models, that sometimes can be classified as CA, have generally the particularity of being parsimonious in their structure.

We study in this paper a morphogenesis model, at the mesoscopic scale, aimed at being simplistic in its rules and variables, but trying to be accurate in the reproduction of existing patterns. The underlying question is to explore the performance of simple mechanisms in reproducing complex urban patterns. We consider abstract processes, namely aggregation and diffusion, candidates as partially explanatory drivers of urban growth, based on population only, that will be detailed in model rationale below. An important aspect we introduce is the quantitative measure of urban form, based on a combination of morphological indicators, to quantify and compare model outputs and real urban patterns. Our contribution is significant on several points: (i) we compute local morphological characteristics on a large spatial extent (full European Union); (ii) we give significant insights into model behavior through extensive exploration of the parameter space; (iii) we show through calibration that the model is able to reproduce most of existing urban forms across Europe, and that these abstract processes are sufficient to explain urban form alone.

The rest of this paper is organized as follows: we first describe formally the model and the morphological indicators. We then detail values of morphological measures on real data, study the behavior of the model by exploring its parameter space and through a semi-analytical approach to a simplified case, and we describe results of model calibration.

%%%%%%%%%%%%%%%%%%%%%%%%
\section*{Material and Methods}
%%%%%%%%%%%%%%%%%%%%%%%%

%%%%%%%%%%%%%%%%%%%%%%%%
\subsection*{Urban growth model}
%%%%%%%%%%%%%%%%%%%%%%%%

%%%%%%%%%%%%%%%%%%%%%%%%
\paragraph*{Rationale}

Our model is based on widely accepted ideas of diffusion-aggregation processes for Urban Processes. The combination of attraction forces with repulsion, due for example to congestion, already yield a complex outcome that has been shown under some simplifying assumptions to be representative of urban growth processes. A model capturing these processes was introduced by~\cite{batty2006hierarchy}, as a cell-based variation of the DLA model~\cite{batty1991generating}. Indeed, the tension between antagonist aggregation and sprawl mechanisms may be an important process in urban morphogenesis. For example \cite{fujita1996economics} opposes centrifugal forces with centripetal forces in the equilibrium view of urban spatial systems, what is easily transferable to non-equilibrium systems in the framework of self-organized complexity: a urban structure is a far-from-equilibrium system that has been driven to this point by these opposite forces. The two contradictory processes of urban concentration and urban sprawl are captured by the model, what allows to reproduce with a good precision a large number of existing morphologies. We can expect aggregation mechanisms such as preferential attachment to be good candidates in urban growth explanation, as it was shown that the Simon model based on them generates power-laws typical of urban systems (scaling laws for example)~\cite{2016arXiv160806313S}. The question at which scale is it possible and relevant to define and try to simulate urban form is rather open, and will in fact depend on which issues are being tackled. Working in a typical setting of morphogenesis, the processes considered are local and our model must have a resolution at the micro-level. We however want to quantify urban form on consistent urban entities, and will work therefore on spatial extents of order 50~100km. We sum up these two aspects by stating that the model is at the \emph{mesoscopic scale}.

\paragraph*{Formalization}

We formalize now the model and its parameters. The world is a square grid of width $N$, in which each cell is characterized by its population $(P_i(t))_{1\leq i\leq N^2}$. We consider the grid initially empty, i.e. $P_i(0)=0$, but the model can be easily generalized to any initial population distribution. The population distribution is updated in an iterative way. At each time step,

\begin{enumerate}
\item Total population is increased by a fixed number $N_G$ (growth rate). Each population unit is attributed independently to a cell following a preferential attachment such that 
\begin{equation}
\Pb{P_i(t+1)=P_i(t)+1|P(t+1)=P(t)+1}=\frac{(P_i(t)/P(t))^{\alpha}}{\sum(P_i(t)/P(t))^{\alpha}}
\end{equation}
The attribution being uniformly drawn if all population are equal to 0.
\item A fraction $\beta$ of population is diffused to cell neighborhood (8 closest neighbors receiving each the same fraction of the diffused population). This operation is repeated $n_d$ times.
\end{enumerate}

The model stops when total population reaches a fixed parameter $P_m$. To avoid bord effects such as reflecting diffusion waves, border cells diffuse their due proportion outside of the world, implying that the total population at time $t$ is strictly smaller than $N_G\cdot t$.

We summarize model parameters in Table~\ref{tab:parameters}, giving the associated processes and values ranges we use in the simulations. The total population of the area $P_m$ is exogenous, in the sense that it is supposed to depend on macro-scale growth patterns on long times. Growth rate $N_G$ captures both endogenous growth rate and migration balance within the area. The aggregation rate $\alpha$ sets the differences in attraction between cells, what can be understood as an abstract attraction coefficient following a scaling law of population. Finally, the two diffusion parameters are complementary since diffusing with strength $n_d\cdot \beta$ is different of diffusing $n_d$ times with strength $\beta$, the later giving flatter configuration.

% table summarizing parameters
%  $\alpha \in [0.1,4],\beta \in [0,0.5],n_d \in \{1,\ldots , 5\}, N_G \in [500,30000], P_m \in [1e4,1e6]$
%%%%%%%%%%%%%
\begin{table}
\caption{{\bf Summary of model parameters.}}
\begin{tabular}{|l|l|l|l|}
\hline
Parameter & Notation & Process & Range\\ \hline
Total population & $P_m$ & Macro-scale growth & $[1e4,1e6]$\\ \hline
Growth rate & $N_G$ & Meso-scale growth  & $[500,30000]$\\ \hline
Aggregation strength & $\alpha$ & Aggregation & $[0.1,4]$\\ \hline
Diffusion strength & $\beta$ & Diffusion & $[0,0.5]$\\ \hline
Diffusion steps & $n_d$ & Diffusion & $\{1,\ldots , 5\}$\\ \hline
\end{tabular}
\label{tab:parameters}
\end{table}
%%%%%%%%%%%%%%

\paragraph*{Measuring the Urban Form}

As our model is only density-based, we propose to quantify its outputs through spatial morphology, i.e. properties of the spatial distribution of density. At the scale chosen, these will be expected to translate various functional properties of the urban landscape. \cite{guerois2008built} studies the form of European cities using a simple measure of density slopes from the center to the periphery. We need however quantities having a certain level of robustness and invariance. For example, two polycentric cities should be classified as morphologically close whereas a direct comparison of distributions (with the Earth Mover Distance for example) could give a very high distance between configurations depending on center positions. The use of fractal indexes is a possibility suggested by~\cite{2016arXiv160808839C}. We choose to refer to the literature in Urban Morphology which proposes an extensive set of indicators to describe urban form~\cite{tsai2005quantifying}. The number of dimensions can be reduced to obtain a robust description with a few number of independent indicators~\cite{Schwarz201029}. Note that here we consider indicators on population density only, and that more elaborated considerations on Urban Form include for example the distribution of economic opportunities and the combination of these two fields through accessibility measures. For the choice of indicators, we follow the analysis done in~\cite{le2015forme} in which a morphological typology of large european cities is obtained.

We give now the formal definition of morphological indicators. Let write $M=N^2$ the number of cells, $d_{ij}$ the distance between cells $i,j$, and $P=\sum_{i=1}^{N} P_i$ total population. We measure Urban Form using:

\begin{enumerate}
\item Rank-size slope $\gamma$, expressing the degree of hierarchy in the distribution, computed by fitting with Ordinary Least Squares a power law distribution by $\ln \left( P_{\tilde{i}}/P_0\right) \sim k + \gamma\cdot \ln \left(\tilde{i}/i_0\right)$ where $\tilde{i}$ are the indexes of the distribution sorted in decreasing order. It is always negative, and values close to zero mean a flat distribution.
\item Entropy of the distribution, that expresses how uniform the distribution is:
\begin{equation}
\mathcal{E} = \sum_{i=1}^{M}\frac{P_i}{P}\cdot \ln{\frac{P_i}{P}}
\end{equation}
$\mathcal{E}=0$ means that all the population is in one cell whereas $\mathcal{E}=0$ means that the population is uniformly distributed.
\item Spatial-autocorrelation given by Moran index, with simple spatial weights given by $w_{ij} = 1/d_{ij}$
\[
I = \frac{\sum_{i\neq j} w_{ij} \left(P_i - \bar{P}\right)\cdot\left(P_j - \bar{P}\right)}{\sum_{i\neq j} w_{ij} \sum_{i}{\left( P_i - \bar{P}\right)}^2}
\]
Positive values will imply aggregation spots (``density centers''), negative values strong local variations whereas $I=0$ corresponds to totally random population values.
\item Mean distance between individuals, which captures population concentration
\[
\bar{d} = \frac{1}{d_M}\cdot \sum_{i<j} \frac{P_i P_j}{P^2} \cdot d_{ij}
\]
where $d_M$ is a normalisation constant taken as the diagonal of the world in our case.
\end{enumerate}

%%%%%%%%%%%%%%%%%%%%%%%%
\begin{figure}[h!]
\centering
\includegraphics[width=0.8\textwidth]{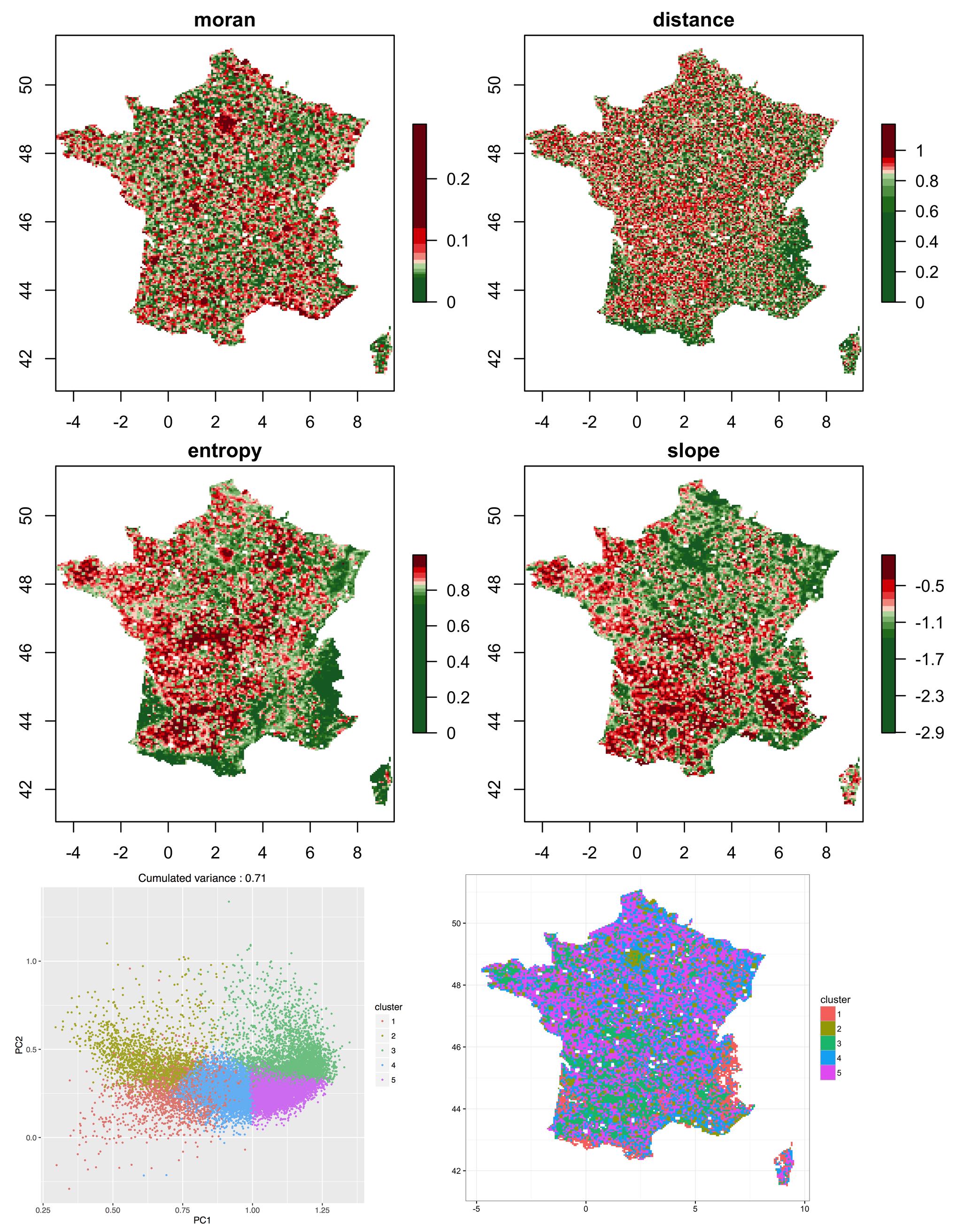}
\caption{\textbf{Empirical values of morphological indicators.} \textit{(Top four maps)} Spatial distribution of the morphological indicators for France. Scale color discretization is done using quantiles to ease map readability. \textit{Bottom Left} Projection of morphological values on the two first components on a Principal Component analysis. Color gives cluster in an unsupervised classification (see text). \textit{Bottom right} Spatial distribution of clusters.}
\label{fig:empirical}
\end{figure}
%%%%%%%%%%%%%%%%%%%%%%%%

%%%%%%%%%%%%%%%%%%%%%%%%
\subsection*{Real Data}

We compute the morphological measures given above on real urban density data, using the population density grid of the European Union at 100m resolution provided openly by Eurostat~\cite{eurostat}. The choice of the resolution, the spatial range, and the shape of the window on which indicators are computed, is made according to the thematic specifications of the model. We consider 50km wide square windows to be in accordance with the expected spatial range of one model instance. As it also does not make sense to have a too detailed resolution because of data quality, we take $N=100$ and aggregate the initial raster data at a 500m resolution to meet this size on real windows. To have a rather continuous distribution of indicators in space, we overlap windows by setting an offset of 10km between each, what also somehow rules out the question of window shape bias by the ``continuity'' of values. We tested the sensitivity to window size by computing samples with 30km and 100km window sizes and obtained rather similar spatial distributions. We show in Fig.~\ref{fig:empirical} maps giving values of indicators for France only to ease maps readability. The first striking feature is the diversity of morphological patterns across the full territory. The auto-correlation is naturally high in Metropolitan areas, with the Parisian surroundings clearly detached. When looking at other indicators, it is interesting to denote regional regimes: rural areas have much less hierarchy in the South than in the North, whereas the average distance is rather uniformly distributed except for mountain areas. Regions of very high entropy are observed in the Center and South-West. To have a better insight into morphological regimes, we use unsupervised classification with a simple k-means algorithm, for which the number of clusters $k=5$ witnesses a transition in inter-cluster variance. The split between classes is plotted in Fig.~\ref{fig:empirical}, bottom-left panel, where we show measures projected on the two first components of a Principal Component Analysis (explaining 71\% of variance). The map of morphological classes confirms a North-South opposition in a background rural regime (clear green against blue), the existence of mountainous (red) and metropolitan (dark green) regimes. Such a variety of settlements forms will be the target for the model.

%%%%%%%%%%%%%%%%%%%%%%%%
\section*{Results}
%%%%%%%%%%%%%%%%%%%%%%%%

%%%%%%%%%%%%%%%%%%%%%%%%
\subsection*{Generation of urban patterns}

\paragraph{Implementation}

The model is implemented both in NetLogo~\cite{wilensky1999netlogo} for exploration and visualization purposes, and in \texttt{Scala} for performance reasons and easy integration into OpenMole~\cite{reuillon2013openmole}, which allows a transparent access to High Performance Computing environments. Computation of indicator values on geographical data is done in \texttt{R} using the \texttt{raster} package~\cite{hijmans2015geographic}. Source code and results are available on the open repository of the project at \texttt{https://github.com/JusteRaimbault/Density}. Raw datasets for real indicator values and simulation results are available on Dataverse at \texttt{http://dx.doi.org/10.7910/DVN/WSUSBA}.

\paragraph{Generated Shapes}

\begin{figure}[h!]
\centering
\includegraphics[width=0.8\textwidth]{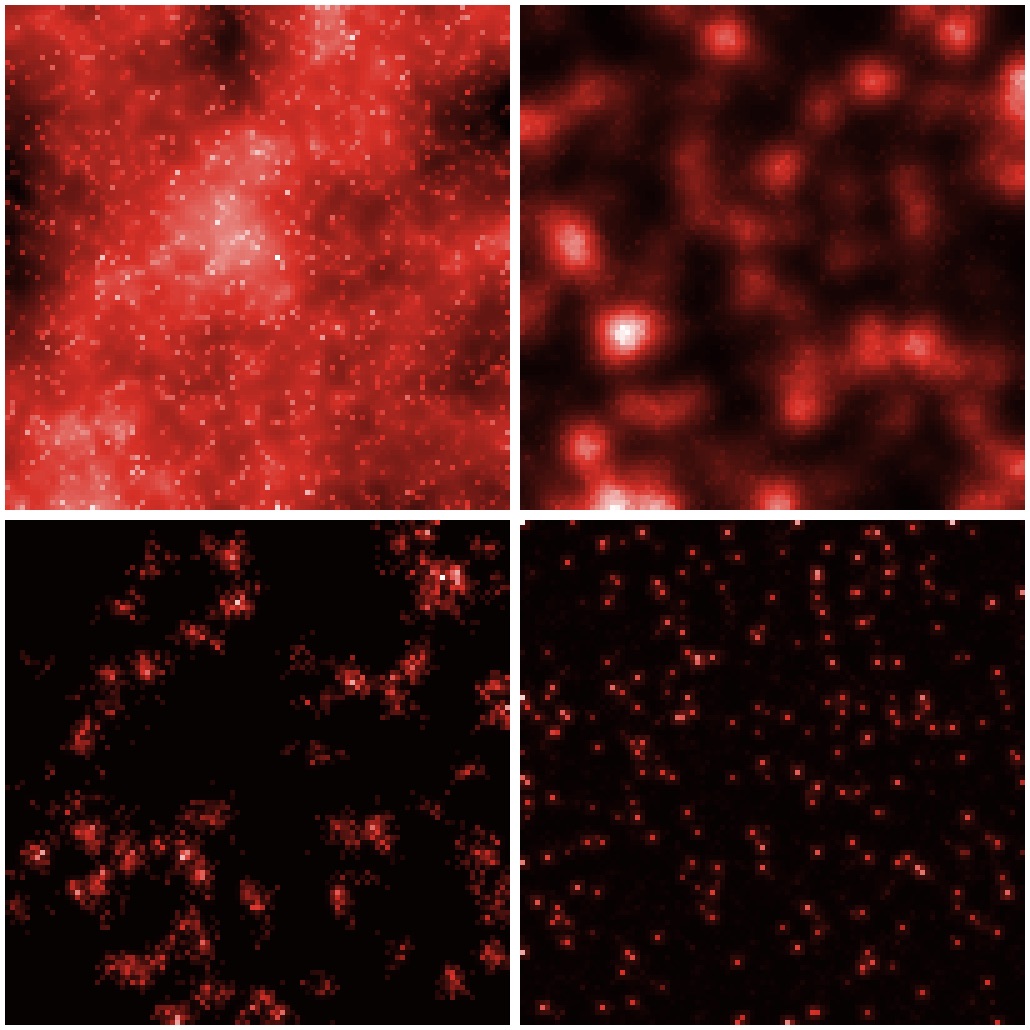}
%ex_sp-diffusion=0.05_sp-growth-rate=76_sp-diffusion-steps=2_sp-alpha-localization=0.4_ticks=995_sp-population=75620.00000000015.png
%ex_sp-diffusion=0.047_sp-growth-rate=274_sp-diffusion-steps=2_sp-alpha-localization=1.4_ticks=197_sp-population=53977.999999999935.png
%ex_sp-diffusion=0.0060_sp-growth-rate=25_sp-diffusion-steps=1_sp-alpha-localization=0.4_ticks=176_sp-population=4400.000000000003.png
%ex_sp-diffusion=0.0060_sp-growth-rate=268_sp-diffusion-steps=1_sp-alpha-localization=1.6_ticks=285_sp-population=76376.00000000033.png
\caption{\textbf{Example of the variety of generated urban shapes.} \textit{(Top left)} Very diffuse urban configuration, $\alpha = 0.4,\beta = 0.05, n_d = 2, N_G = 76, P_m = 75620$; \textit{(Top Right)} Semi-stationary polycentric urban configuration, $\alpha = 1.4,\beta = 0.047, n_d = 2, N_G = 274, P_m = 53977$; \textit{(Bottom Left)} Intermediate settlements (peri-urban or densely populated rural area), $\alpha = 0.4,\beta = 0.006, n_d = 1, N_G = 25, P_m = 4400$; \textit{(Bottom Right)} Rural area, $\alpha = 1.6,\beta = 0.006, n_d = 1, N_G = 268, P_m = 76376$.}
\label{fig:fig2}
\end{figure}

The model has a relatively small number of parameters but is able to generate a large variety of shapes, extending beyond existing forms. In particular, its dynamical nature allows through the interplay between $P_m$ and $N_G$ parameter to choose between final configurations that can be non-stationary or semi-stationary, whereas the interplay between $\alpha$ and $\beta$ modulates the sprawl and the compactness of forms. We run the model for parameters varying in ranges given in Table~\ref{tab:parameters}, for a world size $N=100$. Fig.~\ref{fig:fig2} shows examples of the variety of generated shapes for different parameter values, with corresponding interpretations. The four very different shapes can be obtained with variation of a single parameter sometimes: going from a peri-urban area from a rural area implies an increased aggregation at the same level of diffusion. Note that the model is density driven, and that the parameter $P_m/N_G$ is what really influences the dynamics: the values of $P_m$ are in some cases not directly corresponding to the interpretations we made (for the rural in particular) that are done on densities. A rescaling keeps the settlement form and solves this issue. These examples show the potentiality of the model to produce diverse shapes. We need then to systematically tackle its stochasticity and explore its parameter space.

%%%%%%%%%%%%%
%%%%%%%%%%%%%

%%%%%%%%%%%%%%%%%%%%%%%%
\subsection*{Model Behavior}

In the study of such a computational model of simulation, the lack of analytical tractability must be compensated by an extensive knowledge of model behavior in the parameter space~\cite{banos2013pour}. This type of approach is typical of what Arthur calls the \emph{Computational shift in modern science}~\cite{arthur2015complexity}: knowledge is less extracted through analytical exact resolution than through intensive computational experiments, even for ``simple'' models such as the one we study.

\paragraph{Convergence}

First of all we need to assess the convergence of the model and its behavior regarding stochasticity. We run for a sparse grid of the parameter space consisting of 81 points, with 100 repetitions for each point. Corresponding histograms are shown in~\nameref{S1_Text}. Indicators show good convergence properties: most of indicators are easily statistically discernable across parameter points, and these are distinguished without ambiguity when taking into account all indicators. We use this experiment to find a reasonable number of repetitions needed in larger experiments. For each point, we estimate the Sharpe ratios for each indicators, i.e. mean normalized by standard deviation. The more variable indicator is Moran with a minimal Sharpe ratio of 0.93, but for which the first quartile is at 6.89. Other indicators have very high minimal values, all above 2. Its means than confidence intervals large as $1.5 \cdot \sigma$ are enough to differentiate between two different configurations. In the case of gaussian distribution, we know that the size of the 95\% confidence around the average is given by $2\cdot \sigma \cdot 1.96 / \sqrt{n}$, what gives $1.26 \cdot \sigma$ for $n=10$. We run therefore this number of repetitions for each parameter point in the following, what is highly enough to have statistically significant differences between average as shown above. In the following, when referring to indicator values for the simulated model, we consider the ensemble averages on these stochastic runs.

\paragraph{Exploration of parameter space}

\begin{figure}[h!]
\centering
\includegraphics[width=0.8\textwidth]{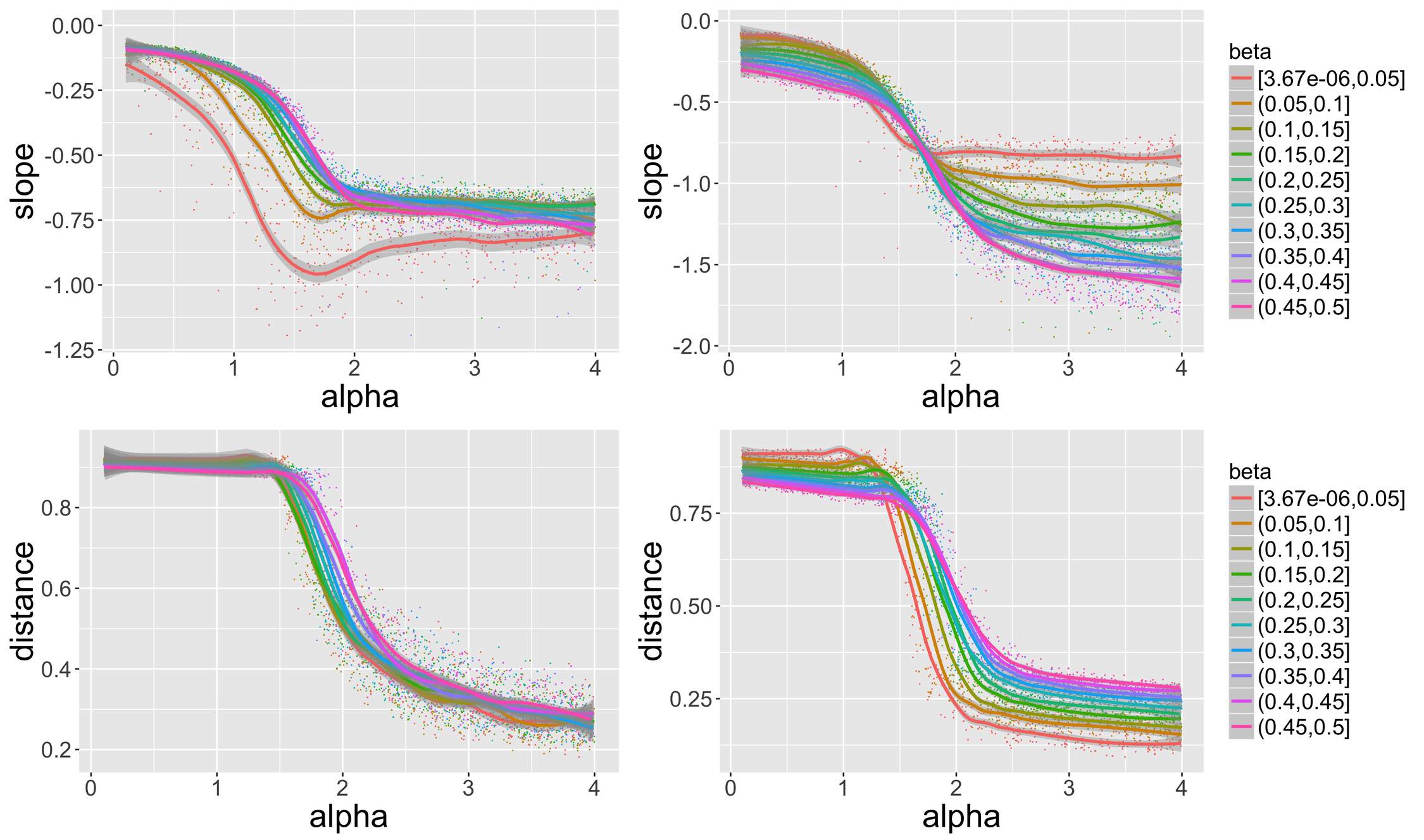}
\caption{\textbf{Behavior of indicators.} Slope $\gamma$ (top row) and average distance $\bar{d}$ (bottom row) as a function of $\alpha$, for different bins for $\beta$ given by curve color, for particular values $n_d=1,P_m/N_G\in\left[13,26\right]$ (left column) and $n_d=4,P_m/N_G\in\left[41,78\right]$ (right column).}
\label{fig:fig3}
\end{figure}

We sample the Parameter space using a Latin Hypercube Sampling, with parameter as $\alpha \in [0.1,4],\beta \in [0,0.5],n_d \in \{1,\ldots , 5\}, N_G \in [500,30000], P_m \in [1e4,1e6]$. This type of cribbing is a good compromise to have a reasonable sampling without being subject to the dimensionality curse within normal computation capabilities. We sample around 80000 parameters points, with 10 repetitions each. Full plots of model behavior as a function of parameters are given in~\nameref{S1_Text}. We show in~\ref{fig:fig3} some particularly interesting behavior for slope $\gamma$ and average distance $\bar{d}$. First of all, the overall qualitative behavior depending on aggregation strength, namely that lower alpha giver less hierarchical and more spread configurations, confirms the expected intuitive behavior. The effect of diffusion strength $\beta$ is more difficult to grasp: the effect is inverted for slope between high and low growth rates but not for distance, that shows an inversion when $\alpha$ varies. In the low $N_G$ case, low diffusion creates more sprawled configuration when aggregation is low, but less sprawled when aggregation is high. Furthermore, all indicators show a more or less smooth transition around $\alpha \simeq 1.5$. Slope stabilize over certain values, meaning that the hierarchy cannot be forced more and indeed depends of the diffusion value, at least for low $N_G$ (right column). In general, higher valued for $P_m/N_G$ increase the effect of diffusion what could have been expected. The existence of a minimum for slope at $n_d=1,P_m/N_G\in\left[13,26\right]$ and lowest $\beta$ is unexpected and witnesses a complex interplay between aggregation and diffusion. The emergence of this ``optimal'' regime is associated with shifts of the transition points in other cases: for example, lowest diffusion imply a transition beginning at lower values of $\alpha$ for average distance. This exploration confirms that complex behavior, in the sense of unpredictable emerging forms, occurs in the model: one cannot say in advance the final form given some parameters, without referring to the full exploration of which we give an overview here.

%%%%%%%%%%%%%
%%%%%%%%%%%%%

%%%%%%%%%%%%%%%%%%%%%%%%
\subsection*{Semi-analytical Analysis}\label{subsec:analytical}

\begin{figure}[h!]
\includegraphics[width=\textwidth]{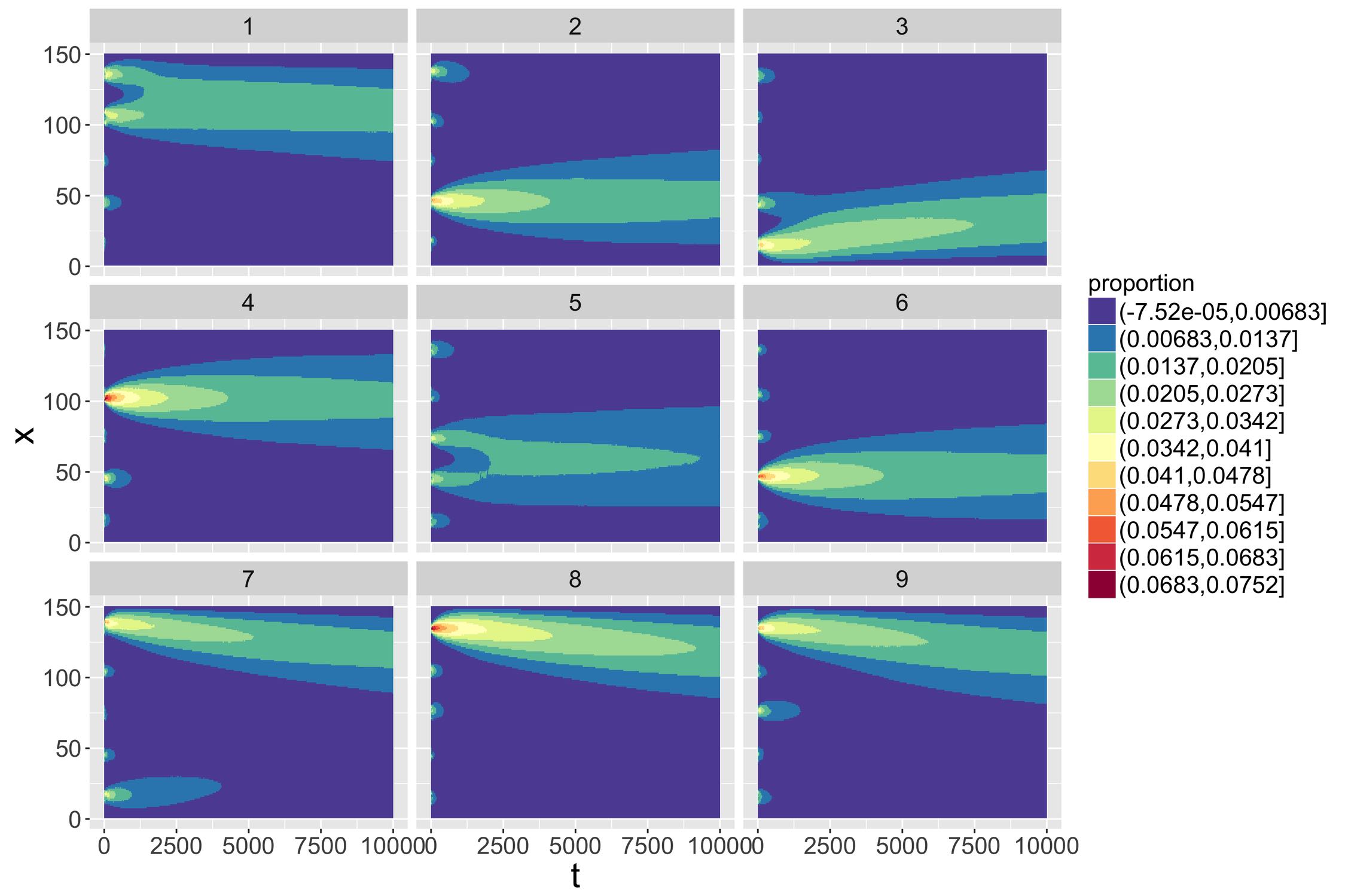}
\caption{\textbf{Randomness and frozen accidents.} We show nine random realizations of the one dimensional system with similar initial conditions, namely five equidistant equally populated initial cells. Parameters are $\alpha = 1.4,\beta =0.1,N_G=10$. Each plot shows time against space, color level giving the proportion of population in each cell.}
\label{fig:fig4}
\end{figure}

Our model can be understood as a type of reaction-diffusion model, that have been widely used in other fields such as biology: similar processes were used for example by Turing in its seminal paper on morphogenesis~\cite{turing1952chemical}. An other way to formulate the model typical to these approaches is by using Partial Differential Equations. We propose to gain insights into long-time dynamics by studying them on a simplified case. We consider the system in one dimension, such that $x\in \left[0;1\right]$ with $1/\delta x$ cells of size $\delta x$. A time step is given by $\delta t$. Each cell is characterized by its population as a random variable $P(x,t)$. We work on their expected values $p(x,t) = \Eb{P(x,t)}$, and assume that $n_d=1$. As developed in Supplementary Material~\nameref{S2_Text}, we show that this simplified process verifies the following PDE:

\begin{equation}\label{eq:pde}
\delta t \cdot \frac{\partial p}{\partial t} = \frac{N_G \cdot p^{\alpha}}{P_{\alpha}(t)} + \frac{\alpha \beta (\alpha - 1) \delta x^2}{2}\cdot \frac{N_G \cdot p^{\alpha-2}}{P_{\alpha}(t)} \cdot \left(\frac{\partial p}{\partial x}\right)^2 + \frac{\beta \delta x^2}{2} \cdot \frac{\partial^2 p}{\partial x^2} \cdot\left[ 1 + \alpha \frac{N_G p^{\alpha - 1}}{P_{\alpha(t)}} \right]
\end{equation}

where $P_{\alpha}(t) = \int_x p(x,t)^{\alpha} dx$. This non-linear equation can not be solved analytically, the presence of integral terms putting it out of standard methods, and numerical resolution must be used~\cite{tadmor2012review}. It is important to note that the simplified model can be expressed by a PDE analog to reaction-diffusion equations. We show in \nameref{S2_Text} that because of the boundaries conditions, density (proportion of population) converges towards a stationary solution at long times, going through intermediate states in which the solution is partially stabilized, in the sense that its evolution speed becomes rather slow. These ``semi-stationary'' states are the ones used in two dimensions along with the dynamical ones. This study confirms that the variety of shapes obtained through the model is permitted both by the interplay of aggregation and diffusion as the equation couples them, but also by the values of $P_m / N_G$ that allow to set the convergence level. Indeed, the sensitivity of the stationary solution to parameters is very low compared to the shape of the world, and using the model in stationary mode would make no sense in our case. Finally, we use this toy case to demonstrate the importance of bifurcations in model dynamics. More precisely, we show that path-dependence is crucial for the final form. As illustrated in Fig.~\ref{fig:fig4}, using an initial condition making the choice ambiguous, corresponding to five equidistant equally populated cells, produces very different trajectories, as generally one of the spots will end dominating the others, but is totally random, witnessing dramatic bifurcations in the system at initial times. This aspect is typically expected in urban systems, and confirms the importance of robust indicators described before.

%%%%%%%%%%%%%
%%%%%%%%%%%%%

%%%%%%%%%%%%%%%
\begin{figure}[h!]
\centering
\includegraphics[width=0.8\textwidth]{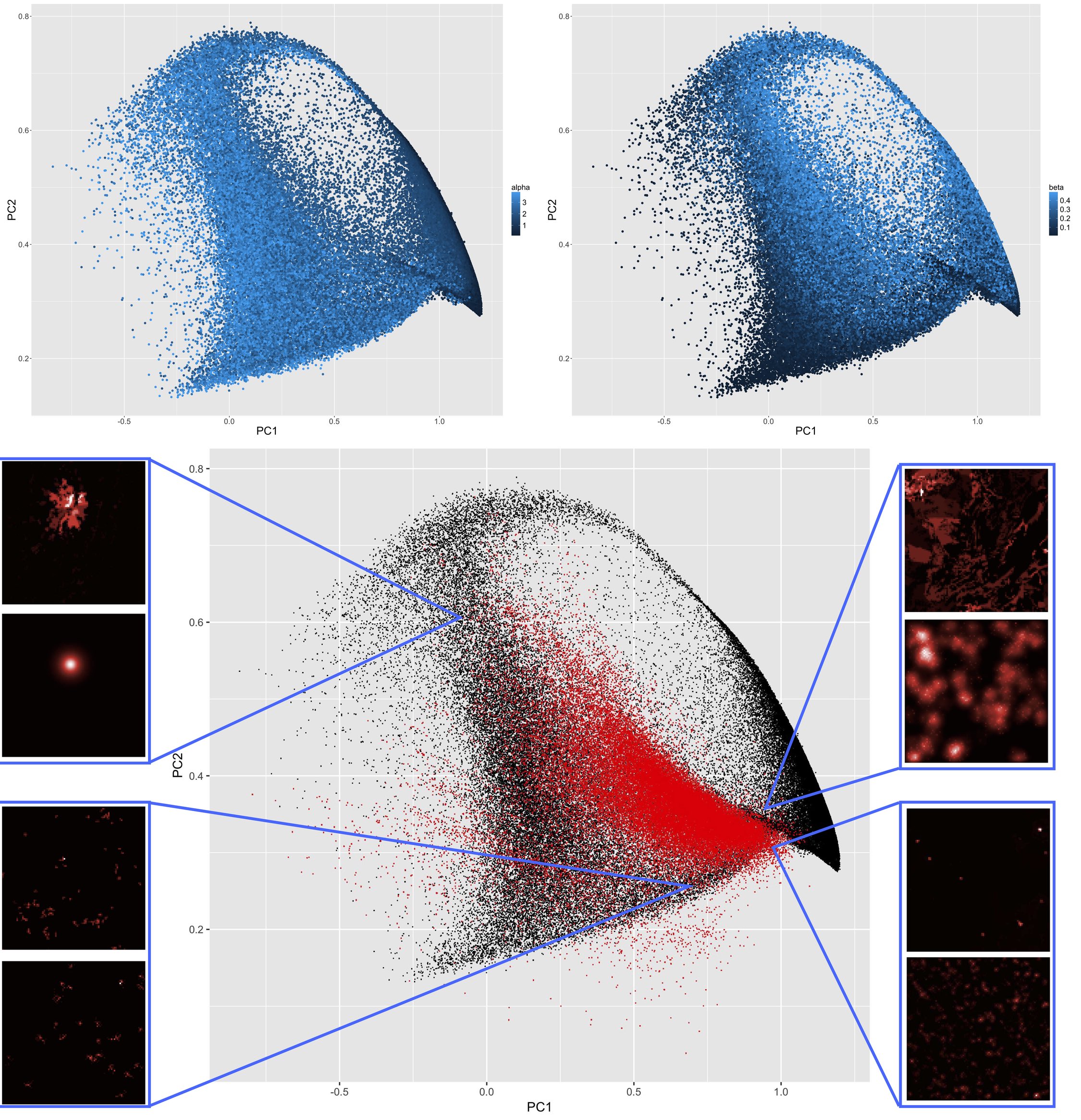}
\caption{\textbf{Model calibration.} \textit{(Top)} Simulated configurations in the two first principal components plan, color level giving the influence of $\alpha$ (left) and of $\beta$ (right); \textit{(Bottom)} Simulated points in the same space (in black) with real configurations (in red). We show around the plot typical examples of real configurations and their simulated counterparts in different regions of the space, the first being the real and the second the simulated in each case: Top left geographical coordinates 25.7361,44.69989 - Romania, Bucharest - simulation parameters $\alpha=3.87,\beta=0.432,N_G=1273,nd=4,P_m=63024$ ; Top right geographical coordinates -2.561874,41.30203 - Spain, Castilla et Leon, Soria - simulation parameters $\alpha=1,\beta=0.166,N_G=100,nd=1,P_m=10017$; Bottom left geographical coordinates 27.16068,65.889 - Finland, Lapland - simulation parameters $\alpha=0.4,\beta=0.006,N_G=25,nd=1,P_m=849$; Bottom right geographical coordinates -2.607152,39.74274 - Spain, Castilla-La Mancha, Cuenca - simulation parameters $\alpha=1.14,\beta=0.108,N_G=637,nd=1,P_m=13235$.}
\label{fig:densitycalib}
\end{figure}
%%%%%%%%%%%%%%

%%%%%%%%%%%%%%%%%%%%%%%%
\subsection*{Model Calibration}

We finally turn to the the calibration of the model, that is done on the morphological objectives. As a single calibration for each real cell is computationally out of reach, we use the previous model exploration and superpose the point clouds with real indicator values. Full scatterplots of all indicators against each other, for simulated and real configurations, are given in~\nameref{S1_Text}. We find that the real point cloud is mostly contained within the simulated, that extend in significantly larger areas. It means that for a large majority of real configuration, there exist model parameters producing in average exactly the same morphological configuration. The highest discrepancy is for the distance indicator, the model failing to reproduce configuration with high distance, low Moran and intermediate hierarchy. These could for example correspond to polycentric configurations with many consequent centers. We consider a more loose calibration constraint, by doing a Principal Component Analysis on synthetic and real morphological values, and consider the two first components only. These represent 85\% of cumulated variance. The rotated point clouds along these dimensions are shown in Fig.~\ref{fig:densitycalib}. Most of real point cloud falls in the simulated one in this simplified setting. We illustrate particular points with real configurations and their simulated counterparts: for example Bucharest, Romania, corresponds to a monocentric semi-stationary configuration, with very high aggregation but also diffusion and a rather low growth rate. Other examples show less populated areas in Spain and Finland. From the plots giving parameter influence, we can show that most real situation fall in the region with intermediate $\alpha$ but quite varying $\beta$. It is consistent with real scaling urban exponents having a variation range rather small (between 0.8 and 1.3 generally~\cite{pumain2006evolutionary}) compared to the one we allowed in the simulations, whereas the diffusion processes may be much more diverse. This way, we have shown that the model is able to reproduce most of existing urban density configuration in Europe, despite its rather simplicity. It confirms that in terms of urban form, most of drivers at this scale can be translated into these abstract processes of aggregation and diffusion, but also that function must be quite correlated with form since the dimension of function (with an additional economic dimension in form for example) is not taken into account in the model.

% Calib 1 :
% real : PC1 = 0.9544645 ; PC2 = 0.3296364 ; coordinates : -2.607152,39.74274 - Spain, Castilla-La Mancha, Cuenca ; synth : PC1 = 0.9487267 PC2 = 0.3245882 ; beta=0.108 ; NG=637 ; nd=1 ; alpha=1.14 ; population=13235.648362769914
% Calib 2 :
% real : PC1 = 0.7004772 ; PC2 = 0.2195029 ; coordinates : 27.16068,65.889 - Finland, Lapland;  synth :  PC1 = 0.6870686 ; PC2 = 0.2287785 ; beta=0.0060 ; NG=25 ; nd=1 ; alpha=0.4 ; population=849.895449367323
% Calib 3 :
% real : PC1 = 1.017064 ; PC2 = 0.3510089 ; coordinates : -2.561874,41.30203 - Spain, Castilla et Leon, Soria ; synth : PC1 = 1.005976 PC2 = 0.3950987 ; beta=0.166; NG=100;nd=1;alpha=1;population=10017.238452906771
% Calib 4 :
% real : PC1 = -0.00177 ; PC2 = 0.6006739 ; coordinates : 25.7361,44.69989 - Romania, Bucharest ; synth : PC1 = -0.0543461 PC2 = 0.5798307 ; beta=0.432;NG=1273;nd=4;alpha=3.87;population=63024.359885979036

%%%%%%%%%%%%%%%%%%%%%%%%
\section*{Discussion}

\paragraph{Calibration and model refinement}

\begin{figure}[h!]
\centering
\includegraphics[width=0.8\textwidth]{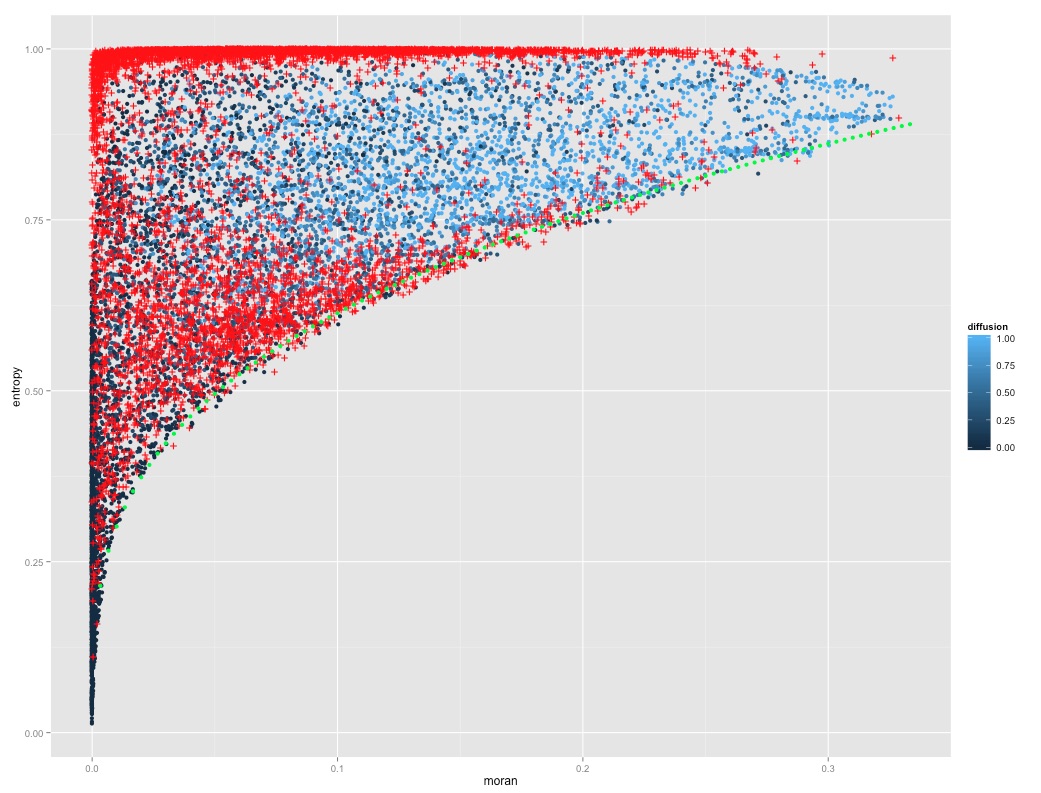}
\caption{\textbf{PSE exploration.} Scatterplots of Moran against Entropy, with blue points obtained with LHS and red with PSE exploration. Green dashed line gives feasible lower bound.}
\label{fig:fig6}
\end{figure}

Further work on this simple model may consist in extracting the exact parameter space covering all real situations and provide interpretation of its shape, in particular through correlations between parameters and expressions of boundaries functions. Its volume in different directions should furthermore give the relative importance of parameters. Concerning the feasible space for the model of simulation itself, we tested a targeted exploration algorithm, giving promising results. More precisely, the Parameter Space Exploration (PSE) algorithm~\cite{10.1371/journal.pone.0138212} which is implemented in OpenMole, is aimed at determining all the possible outputs of a simulation models, i.e. samples its output space rather than input space. We obtain promising results as shown in Fig.~\ref{fig:fig6}: we find that the lower bound in Moran-entropy plan, confirmed by the algorithm, unexpectedly exhibit a scaling relationship. It would mean that at a given level of auto-correlation, that one could want to attain for sustainability reasons for example (optimality through co-location), imposes a minimal disorder in the configuration of activities. Other relations between indicators and as a function of parameters can be the object of similar future developments. The question of doing a dynamical calibration of the model, i.e. trying to reproduce configurations at successive times, is conditioned to the availability of population data at this resolution in time.

% application of Calibration Profile algo to check relative influence of parameters ? out of context not interesting

%%%%%%%%%%%%%
%%%%%%%%%%%%%

We aimed at using abstract processes rather than having a highly realistic model. Tuning some mechanisms is possible to have a model closer to reality in microscopic processes: for example thresholding the local population density, or stopping the diffusion at a given distance from the center if it is well defined. It is however far from clear if these would produce such a variety of forms and could be calibrated in a similar way, as being accurate locally does not mean being accurate at the mesoscopic level for morphological indicators. Allowing the parameters to locally vary, i.e. being non stationary in space, or adding randomness to the diffusion process, are also potential model refinements.

\paragraph{Integration into a multi-scale growth model}

The question of the generic character of the model is also open: would it work as well when trying to reproduce Urban Forms on very different systems such as the United States or China. A first interesting development would be to test it on these systems and at slightly different scales (1km cell for example). Finally, we believe that a significant insight into the non-stationarity of Urban Systems would be allowed by its integration into a multi-scale growth model. Urban growth patterns have been empirically shown to exhibit multi-scale behavior~\cite{zhang2013identifying}. Here at the meso-scale, total population and growth rates are fixed by exogenous conditions of processes occurring at the macro-scale. It is particularly the aim of spatial growth models such as the Favaro-pumain model~\cite{favaro2011gibrat} to determine such parameters through relations between cities as agents. One would condition the morphological development in each area to the values of the parameters determined at the level above. In that setting, one must be careful of the role of the bottom-up feedback: would the emerging urban form influence the macroscopic behavior in its turn ? Such multi-scale complex model are promising but must be considered carefully.

%  determination of effective independent dimensions of the urban system at this scale ?

%%%%%%%%%%%%%%%%%%%%%%%%
\section*{Conclusion}

In conclusion, we have provided a calibrated spatial urban morphogenesis model at the mesoscopic scale that can reproduce any European urban pattern in terms of morphology. We demonstrate that the abstract processes of aggregation and diffusion are sufficient to capture urban growth processes at this scale. It is meaningful in terms of policies based on urban form such as energy efficiency, but also means that issues out of this scope must be tackled at other scales or through other dimensions of urban systems.

%%%%%%%%%%%%%%%%%%%%%%%%
\section*{Supporting Information}

% Include only the SI item label in the subsection heading. Use the \nameref{label} command to cite SI items in the text.

\subsection*{S1 Text}
\label{S1_Text}
{\bf Extended Model Exploration.} Extended figures for model exploration.

\subsection*{S2 Text}
\label{S2_Text}
{\bf Semi-analytical Analysis.} Analytical and numerical developments for the simplified model.

%\subsection*{S2 Text}
%\label{S2_Text}
%{\bf Extended empirical study of morphological indicators.} Contains maps of indicators for all Europe, statistical distributions of indicators, and some sensitivity analysis to grid size and typology parameters.

%%%%%%%%
% Potential supplementary material :
%  - sensitivity to neighborhood
%  - sensitivity of real values to window size

%%%%%%%%%%%%%%%%%%%%%%%%
%\section*{Acknowledgments}

%\section*{References}
% Either type in your references using
% \begin{thebibliography}{}
% \bibitem{}
% Text
% \end{thebibliography}
%
% OR
%
% Compile your BiBTeX database using our plos2015.bst
% style file and paste the contents of your .bbl file
% here.
% 

%\bibliographystyle{plos2015}
%\bibliography{/Users/Juste/Documents/ComplexSystems/CityNetwork/Biblio/Bibtex/CityNetwork,biblio}

\newpage

\section*{S1 Text : Extended Figures for Model Exploration}

\subsection*{Convergence}

\begin{figure}
\centering
\includegraphics[width=0.8\textwidth]{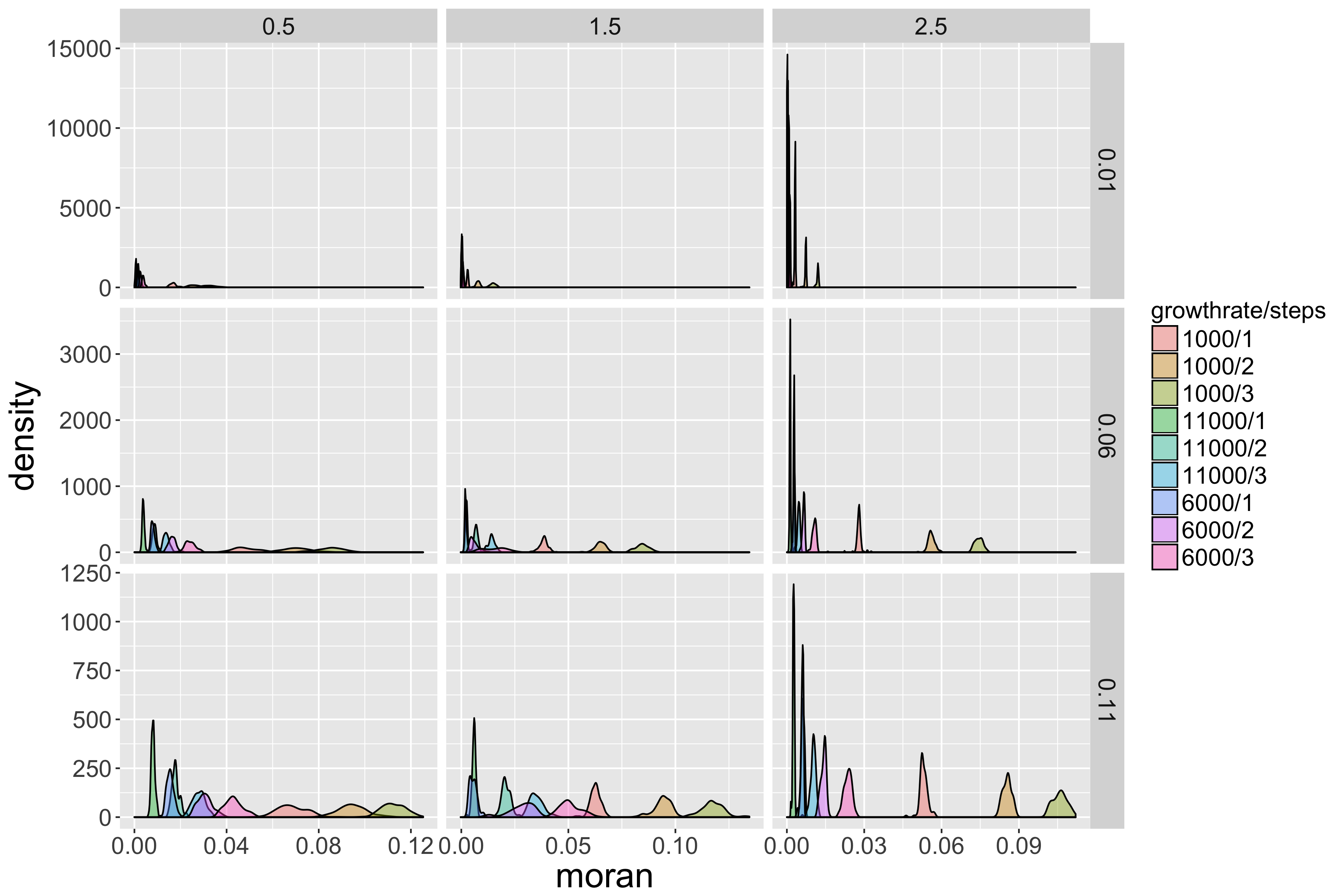}\\
\includegraphics[width=0.8\textwidth]{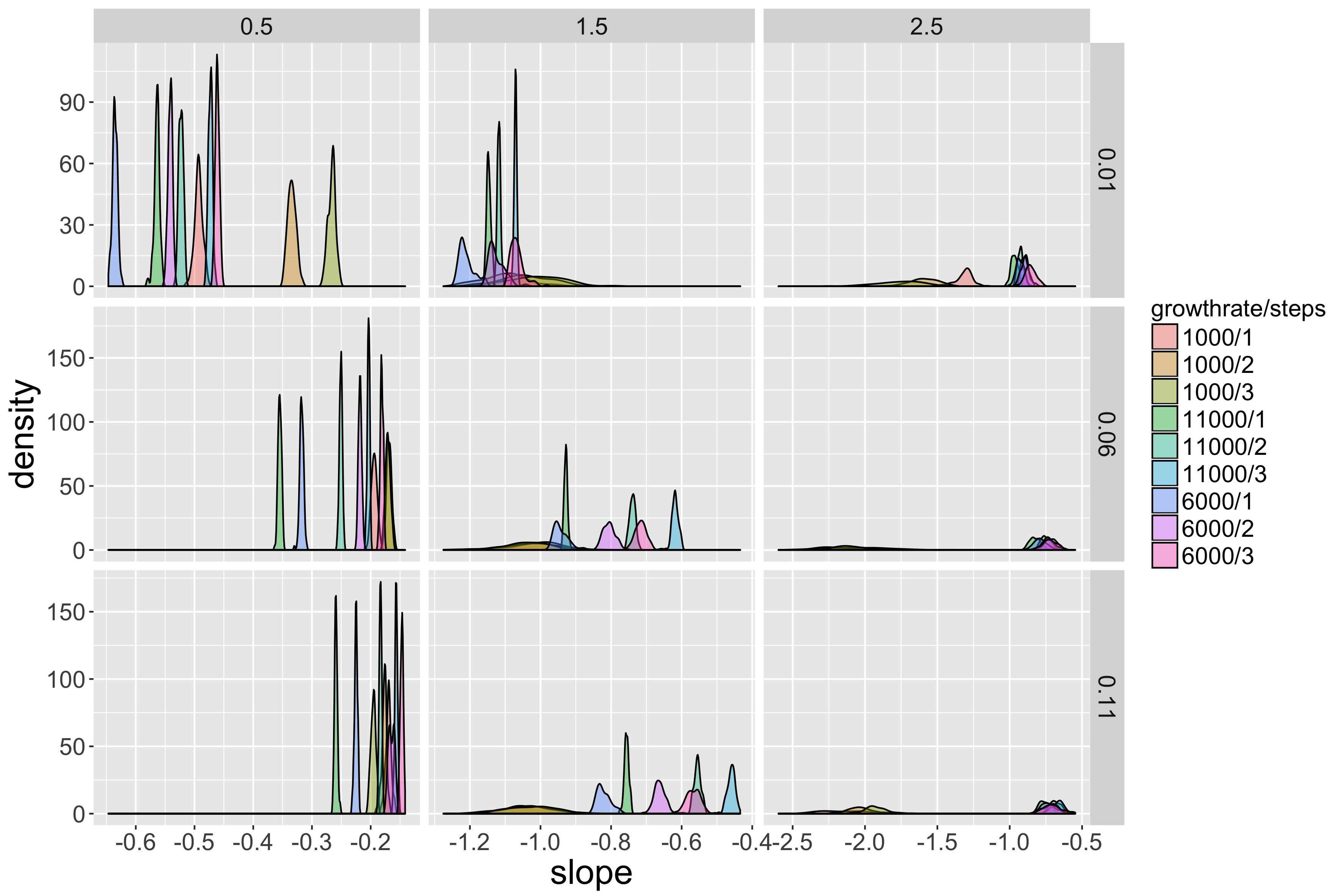}
\caption{Histograms for Moran index (top) and slope (bottom), for varying $\alpha$ (columns), $\beta$ (rows), $N_G$ and $n_d$ (colors).}
\label{fig:histograms}
\end{figure}

Histograms for the 81 parameters points for which we did 100 repetitions are given in Fig.~\ref{fig:histograms}, for Moran index and slope indicators. Other indicators showed similar convergence patterns. The visual exploration of histograms confirms the numerical analysis done in main text for statistical convergence.

% histograms

% and bimodal statistical distribution for cumulated points in the parameter space confirm the existence of superposed regimes : gaussian distribution gives stationary configurations, whereas inverse log-normal distribution are close to real data shape and correspond to non-stationary regime.
% -> not sure it is interesting to look at cumulated histograms.

% Figures :

%%%%%%%%%%%%%%%%%%%%
%%%%%%%%%%%%%%%%%%%%

\subsection*{Indicators Behavior}

% full plots behavior

We show in Fig. to Fig. 10 the full behavior of all indicators, with all parameters varying, obtained through the extensive exploration, from which the plots in main text have been extracted. Because of the complex nature of emergent urban form, one can not predict output values without referring to this ``exhaustive'' parameter sweep.

%%%%%%%%%%%%%%%%%%%%
\begin{figure}
\centering
\includegraphics[width=0.8\textwidth]{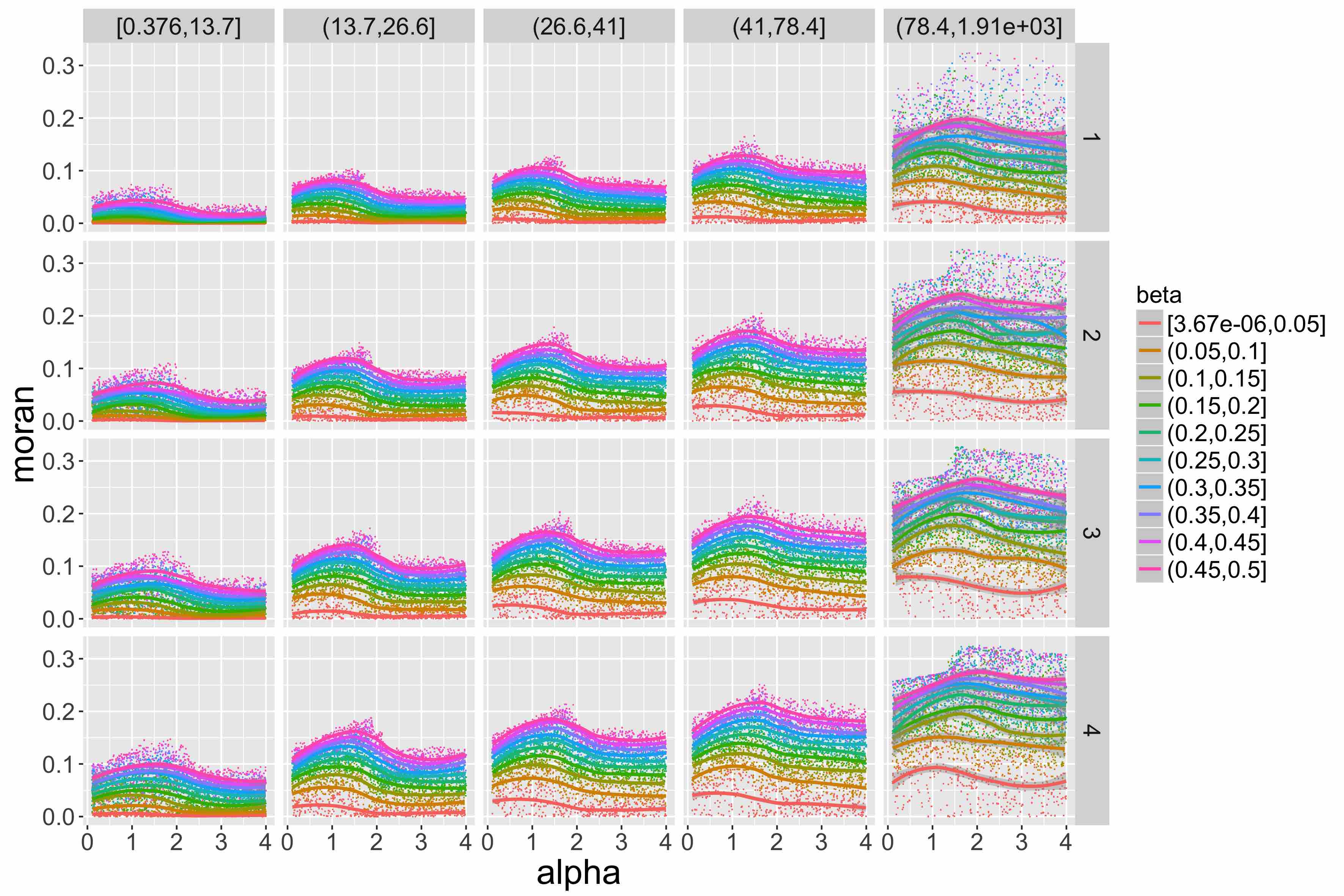}
\includegraphics[width=0.8\textwidth]{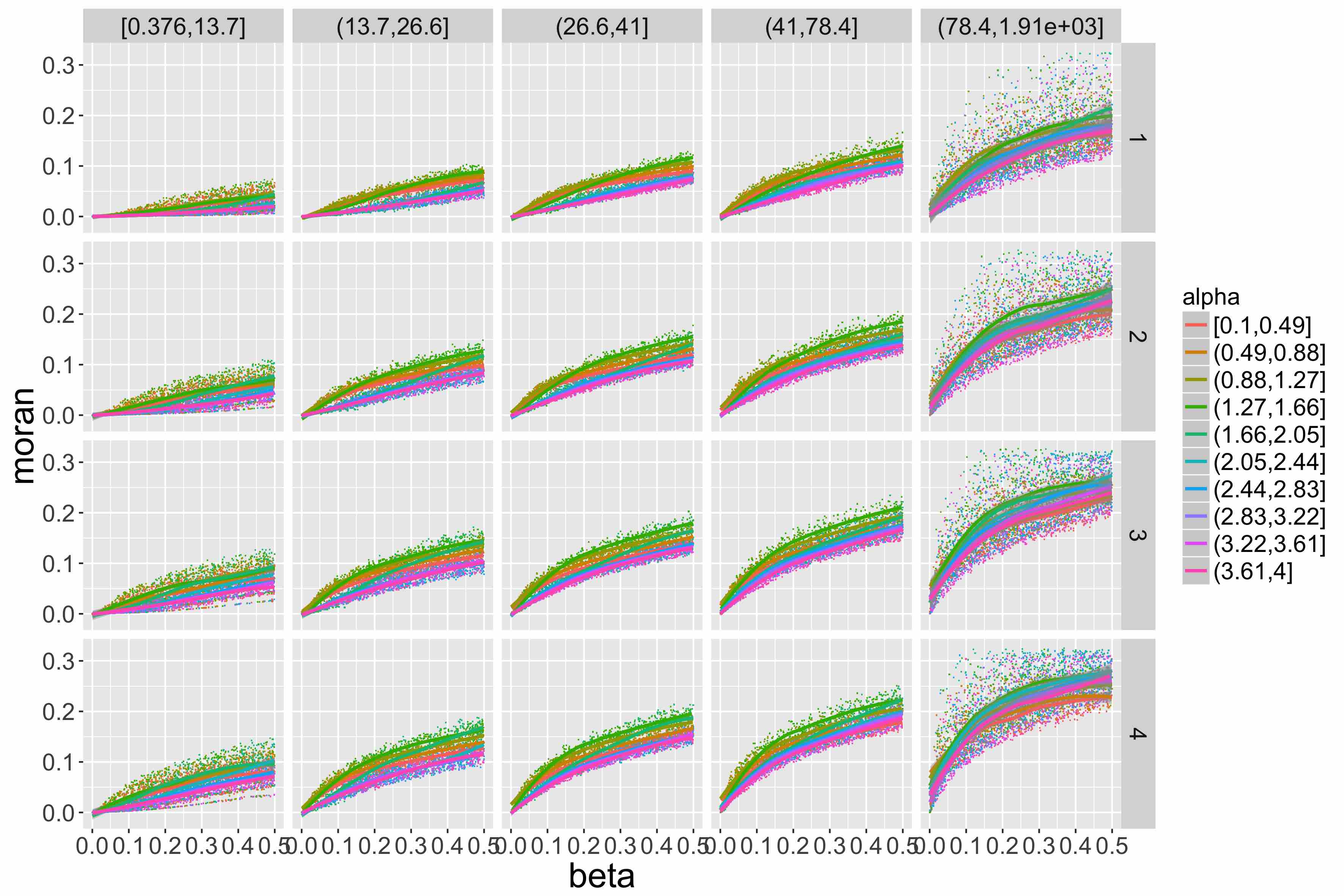}
\caption{Moran index as a function of $\alpha$ (Top) and $\beta$ (Bottom) for varying $\beta$ (resp. $\alpha$) given by color, and varying $n_d$ (rows) and $N_G$ (columns).}
\label{fig:moran}
\end{figure}
%%%%%%%%%%%%%%%%%%%%

%%%%%%%%%%%%%%%%%%%%
\begin{figure}
\centering
\includegraphics[width=0.8\textwidth]{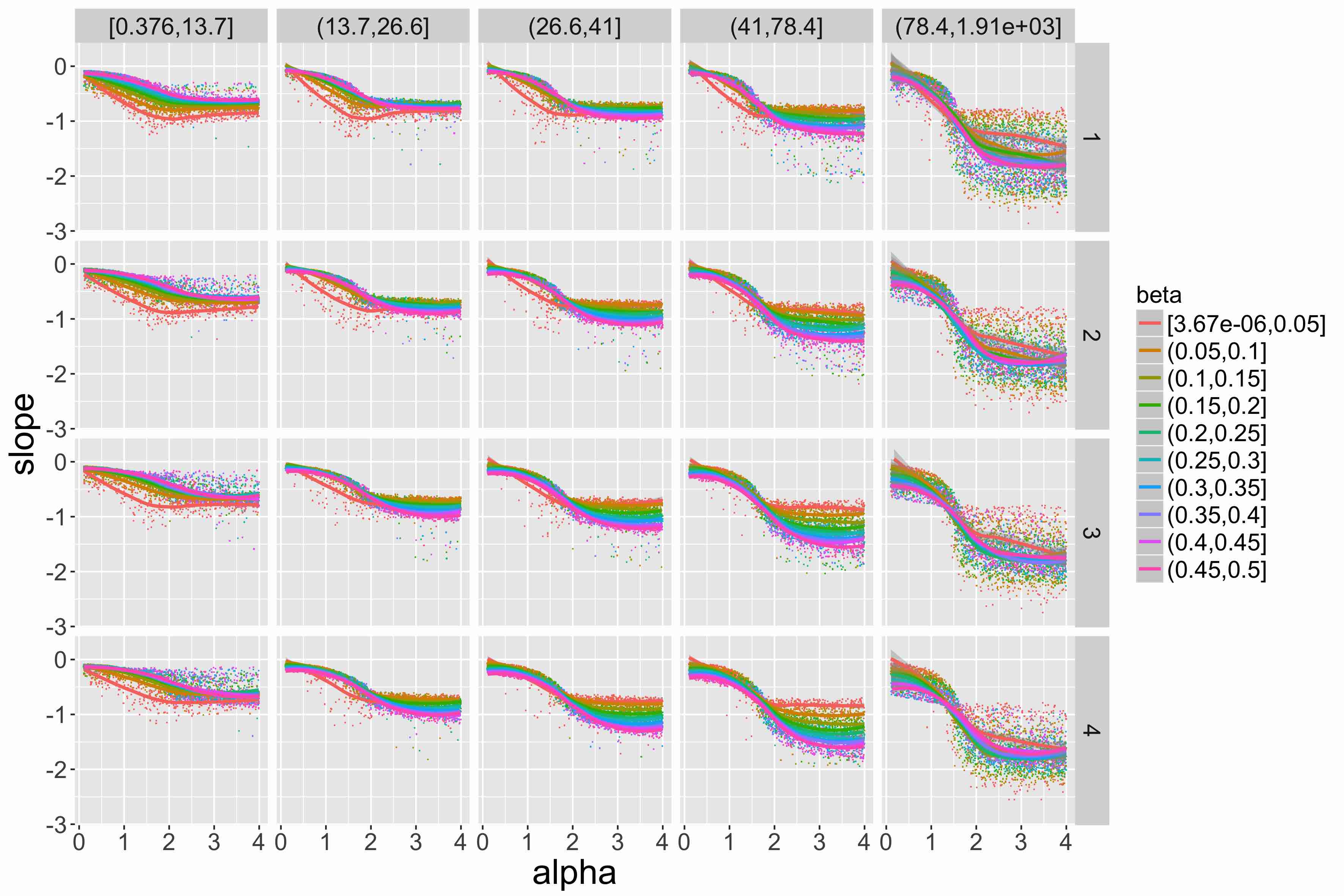}
\includegraphics[width=0.8\textwidth]{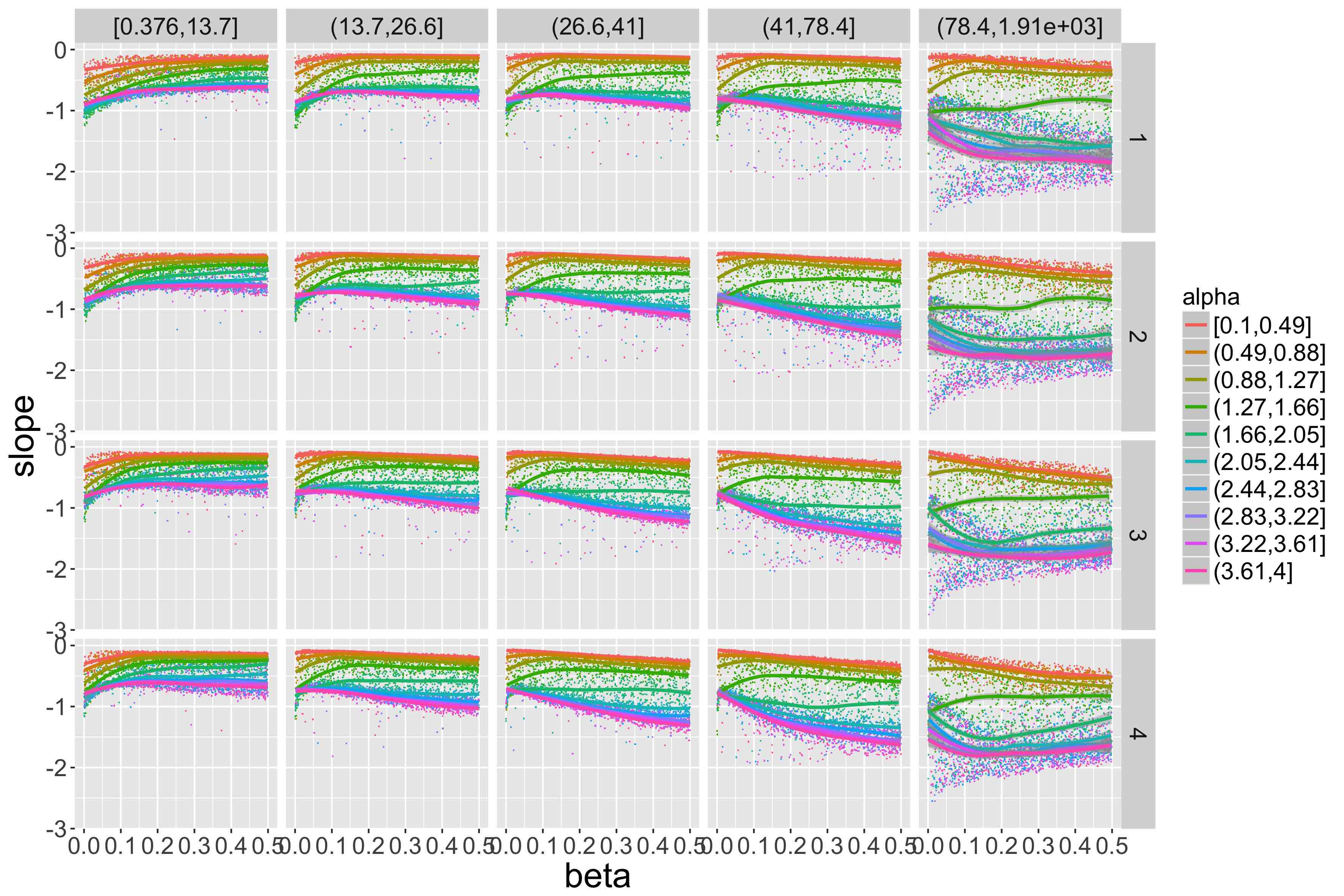}
\caption{Slope as a function of $\alpha$ (Top) and $\beta$ (Bottom) for varying $\beta$ (resp. $\alpha$) given by color, and varying $n_d$ (rows) and $N_G$ (columns).}
\label{fig:slope}
\end{figure}
%%%%%%%%%%%%%%%%%%%%

%%%%%%%%%%%%%%%%%%%%
\begin{figure}
\centering
\includegraphics[width=0.8\textwidth]{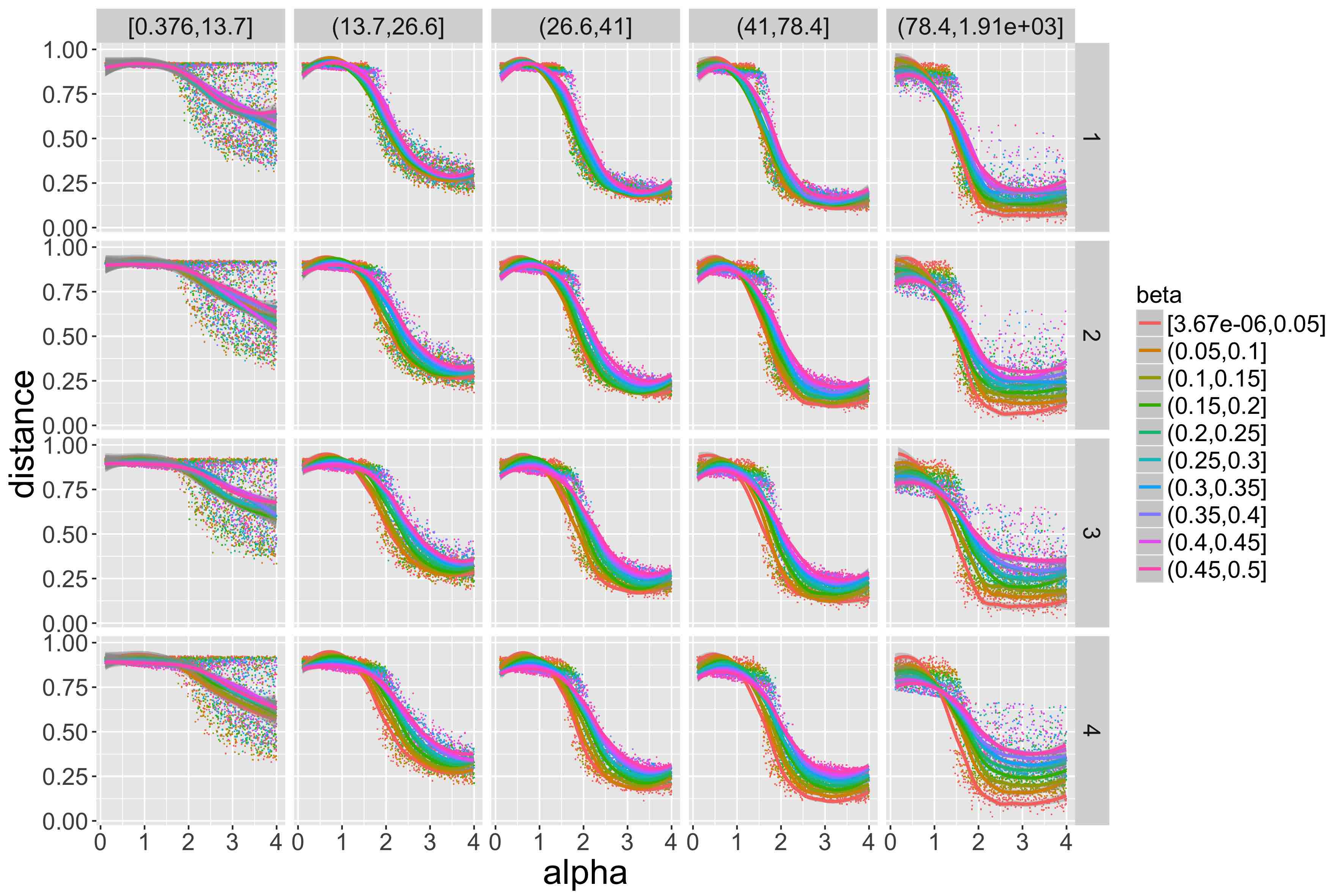}
\includegraphics[width=0.8\textwidth]{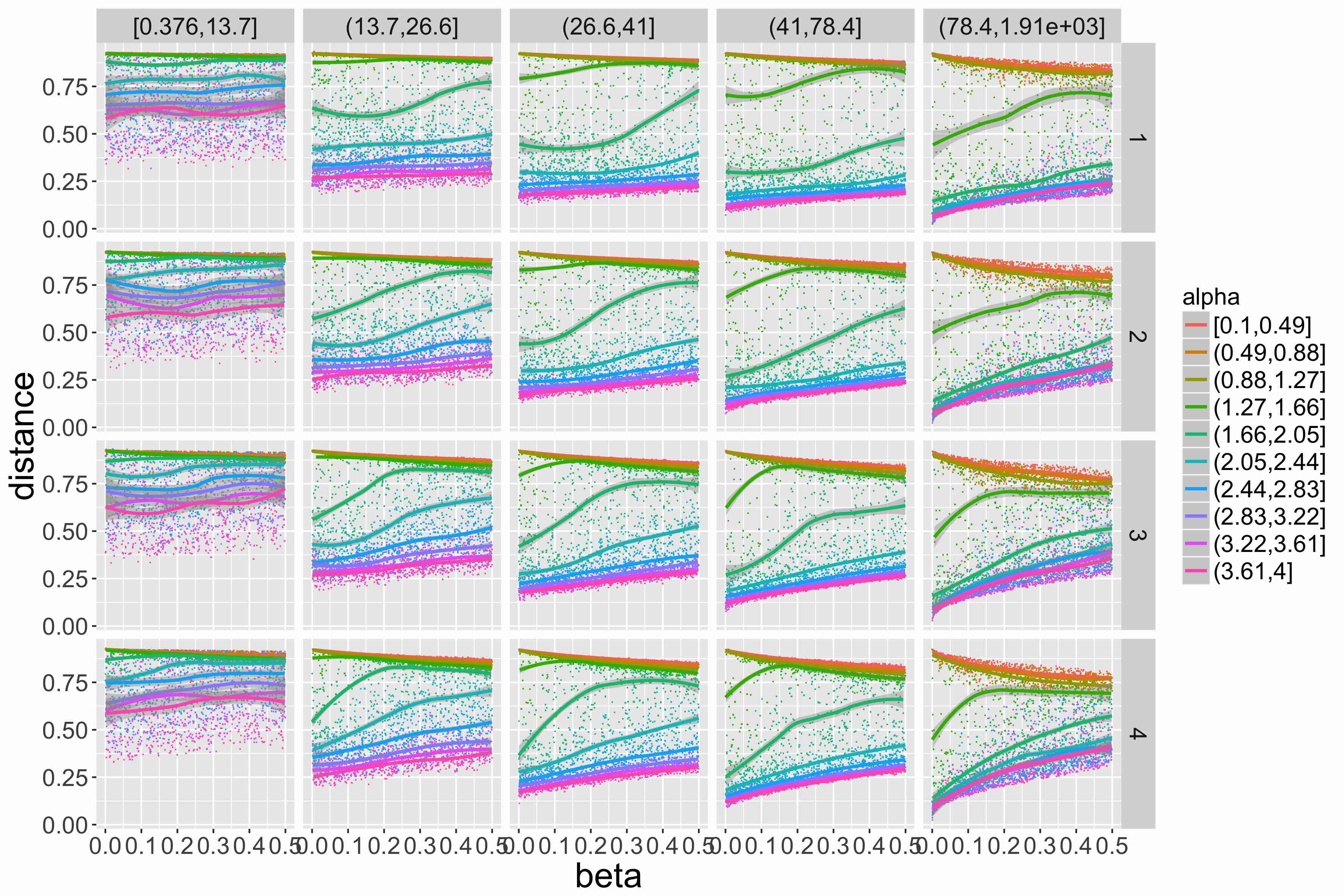}
\caption{Average distance index as a function of $\alpha$ (Top) and $\beta$ (Bottom) for varying $\beta$ (resp. $\alpha$) given by color, and varying $n_d$ (rows) and $N_G$ (columns).}
\label{}
\end{figure}
%%%%%%%%%%%%%%%%%%%%

%%%%%%%%%%%%%%%%%%%%
\begin{figure}
\centering
\includegraphics[width=0.8\textwidth]{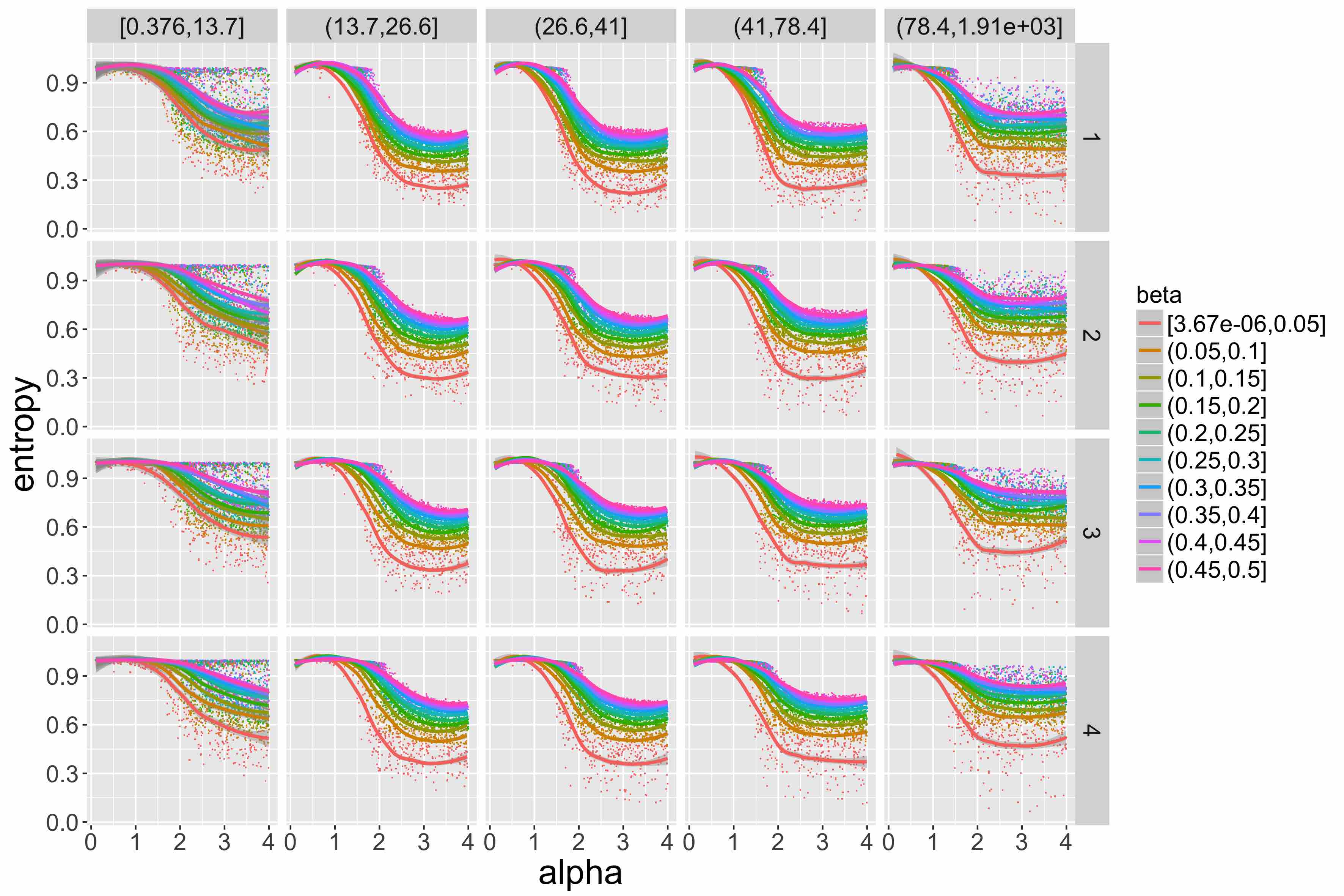}
\includegraphics[width=0.8\textwidth]{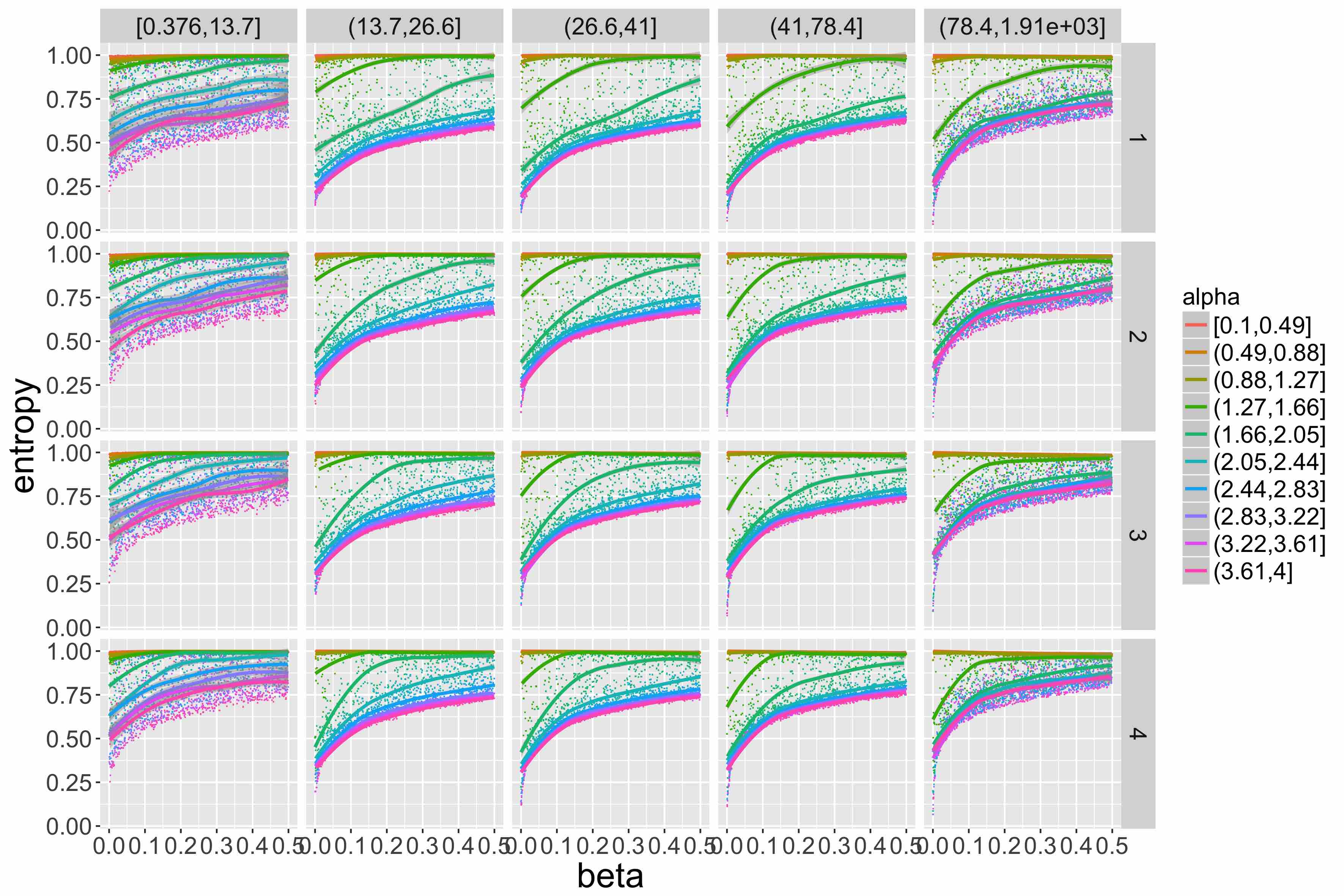}
\caption{Entropy as a function of $\alpha$ (Top) and $\beta$ (Bottom) for varying $\beta$ (resp. $\alpha$) given by color, and varying $n_d$ (rows) and $N_G$ (columns).}
\label{}
\end{figure}
%%%%%%%%%%%%%%%%%%%%

\subsection*{Indicators Scatterplots}

% scatterplots - with real points

\begin{figure}
\includegraphics[width=\textwidth]{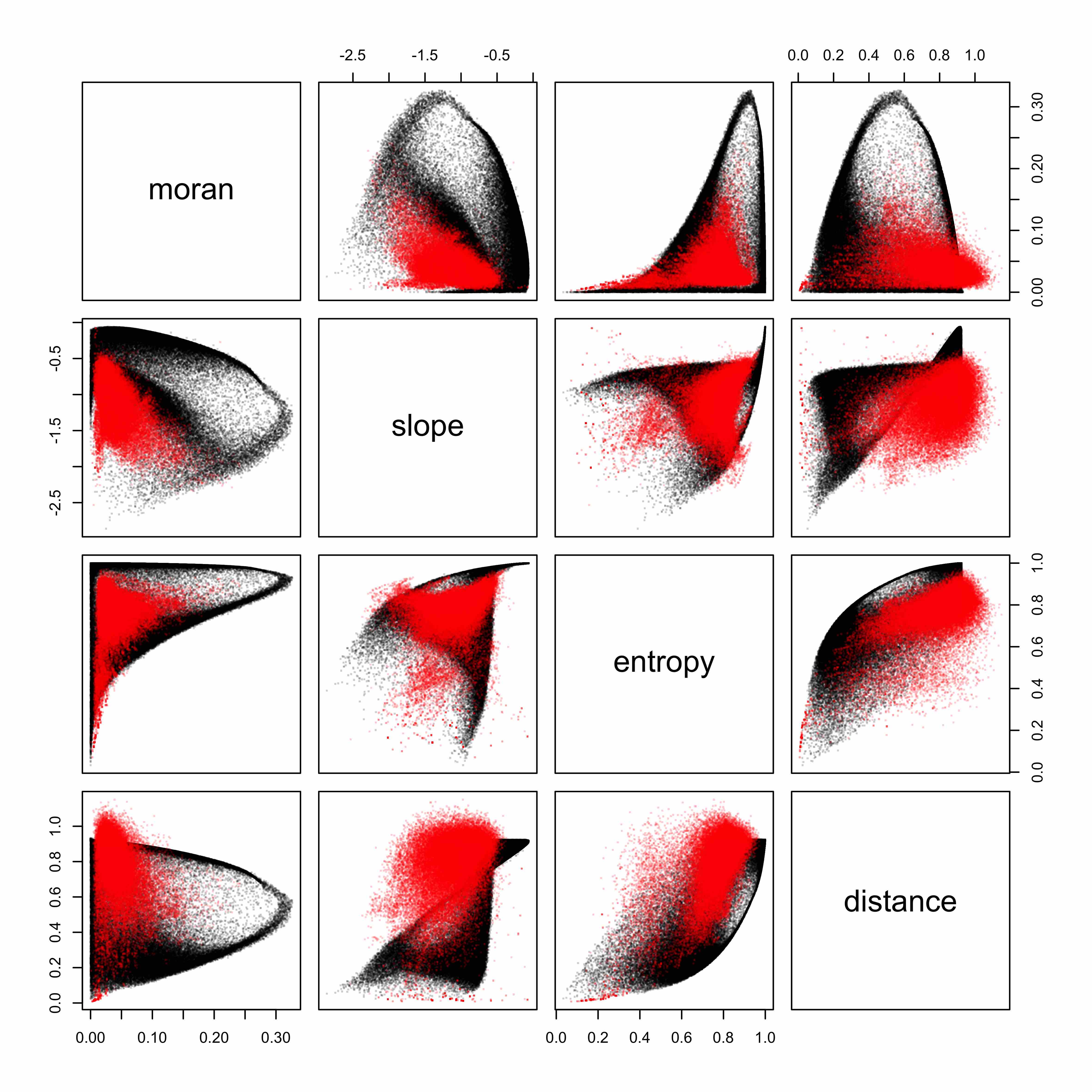}
\caption{Scatterplots of indicators distribution in the sampled hypercube of the parameter space. Red points correspond to real data.}
\label{fig:densityscatter}
\end{figure}

We show finally the full scatterplots of indicators, with real data points, in Fig.~\ref{fig:densityscatter}. These are preliminary step of the calibration on principal components, and we can see on these on which dimensions the model fails relatively to fit real data (in particular average distance).

%%%%%%%%%%%%%
%%%%%%%%%%%%%

\newpage

\section*{S2 Text : Semi-analytical analysis of the simplified model}

\subsection*{Partial Differential Equation}

We propose to derive the PDE in a simplified setting. To recall the configuration given in main text, the system has one dimension, such that $x\in \mathbb{R}$ with $1/\delta x$ cells of size $\delta x$, and we use the expected values of cell population $p(x,t) = \Eb{P(x,t)}$. We furthermore take $n_d=1$. Larger values would imply derivatives at an order higher than 2 but the following results on the existence of a stationary solution should still hold. 

Denoting $\tilde{p}(x,t)$ the intermediate populations obtained after the aggregation stage, we have

\[
\tilde{p}(x,t) = p(x,t) + N_g\cdot \frac{p(x,t)^{\alpha}}{\sum_x p(x,t)^{\alpha}}
\]

since all populations units are added independently. If $\delta x \ll 1$ then $\sum_x p^{\alpha} \simeq \int_x p(x,t)^{\alpha}dx$ and we write this quantity $P_{\alpha}(t)$. We furthermore write $p=p(x,t)$ and $\tilde{p} = \tilde{p}(x,t)$ in the following for readability.

The diffusion step is then deterministic, and for any cell not on the border ($0<x<1$), if $\delta t$ is the interval between two time steps, we have

\[
\begin{split}
p(x,t+\delta t) & = (1 - \beta) \cdot \tilde{p} + \frac{\beta}{2} \left[\tilde{p}(x-\delta x,t) + \tilde{p}(x+\delta x,t)\right]\\
& = \tilde{p} + \frac{\beta}{2} \left[\left(\tilde{p}(x+\delta x,t) - \tilde{p}\right) - \left(\tilde{p} - \tilde{p}(x-\delta x,t)\right)\right]
\end{split}
\]

Assuming the partial derivatives exist, and as $\delta x \ll 1$, we make the approximation $\tilde{p}(x+\delta x,t) - \tilde{p} \simeq \delta x\cdot \frac{\partial \tilde{p}}{\partial{x}}(x,t)$, what gives 

\[
\left(\tilde{p}(x+\delta x,t) - \tilde{p}\right) - \left(\tilde{p} - \tilde{p}(x-\delta x,t)\right) = \delta x \cdot \left(\frac{\partial \tilde{p}}{\partial{x}}(x,t) - \frac{\partial \tilde{p}}{\partial{x}}(x - \delta x,t)\right)
\]

and therefore at the second order

\[
p(x,t+\delta t) = \tilde{p} + \frac{\beta \delta x^2}{2} \cdot \frac{\partial^2 \tilde{p}}{\partial x^2}
\]

Substituting $\tilde{p}$ gives

\[
\begin{split}
\frac{\partial^2 \tilde{p}}{\partial x^2} & = \frac{\partial^2 p}{\partial x^2} + \frac{N_G}{P_\alpha}\cdot \frac{\partial}{\partial x}\left[\alpha \frac{\partial p}{\partial x} p^{\alpha - 1}\right]\\
& = \frac{\partial^2 p}{\partial x^2} + \alpha \frac{N_G}{P_\alpha} \left[\frac{\partial^2 p}{\partial x^2} p^{\alpha - 1} + (\alpha - 1) \left( \frac{\partial p}{\partial x}\right)^2 p^{\alpha - 2}\right]
\end{split}
\]

By supposing that $\frac{\partial p}{\partial t}$ exists and that $\delta t$ is small, we have $p(x,t+\delta t) - p(x,t) \simeq \delta t\frac{\partial p}{\partial t}$, what finally yields , by combining the results above, the partial differential equation

\begin{equation}\label{eq:pde}
\delta t \cdot \frac{\partial p}{\partial t} = \frac{N_G \cdot p^{\alpha}}{P_{\alpha}(t)} + \frac{\alpha \beta (\alpha - 1) \delta x^2}{2}\cdot \frac{N_G \cdot p^{\alpha-2}}{P_{\alpha}(t)} \cdot \left(\frac{\partial p}{\partial x}\right)^2 + \frac{\beta \delta x^2}{2} \cdot \frac{\partial^2 p}{\partial x^2} \cdot\left[ 1 + \alpha \frac{N_G p^{\alpha - 1}}{P_{\alpha(t)}} \right]
\end{equation}

Initial conditions should be specified as $p_0(x) = p(x,t_0)$. To have a well-posed problem similar to more classical PDE problems, we need to assume a domain and boundary conditions. A finite support is expressed by $p(x,t)=0$ for all $t$ and $x$ such that $\left|x\right|>x_m$.

%An infinite domain implies that density, in the sense of population proportion $d(x,t) = \frac{p(x,t)}{P_1(t)}$, goes to zero anywhere when time goes to infinity. Indeed, $P_1(t)=N_G\cdot t$. If $d(x,t)$ does not vanish, there exist $t_1$ such that  -> not that simple indeed

\subsection*{Stationary solution for density}

\begin{figure}[!h]
\centering
\includegraphics[width=\textwidth]{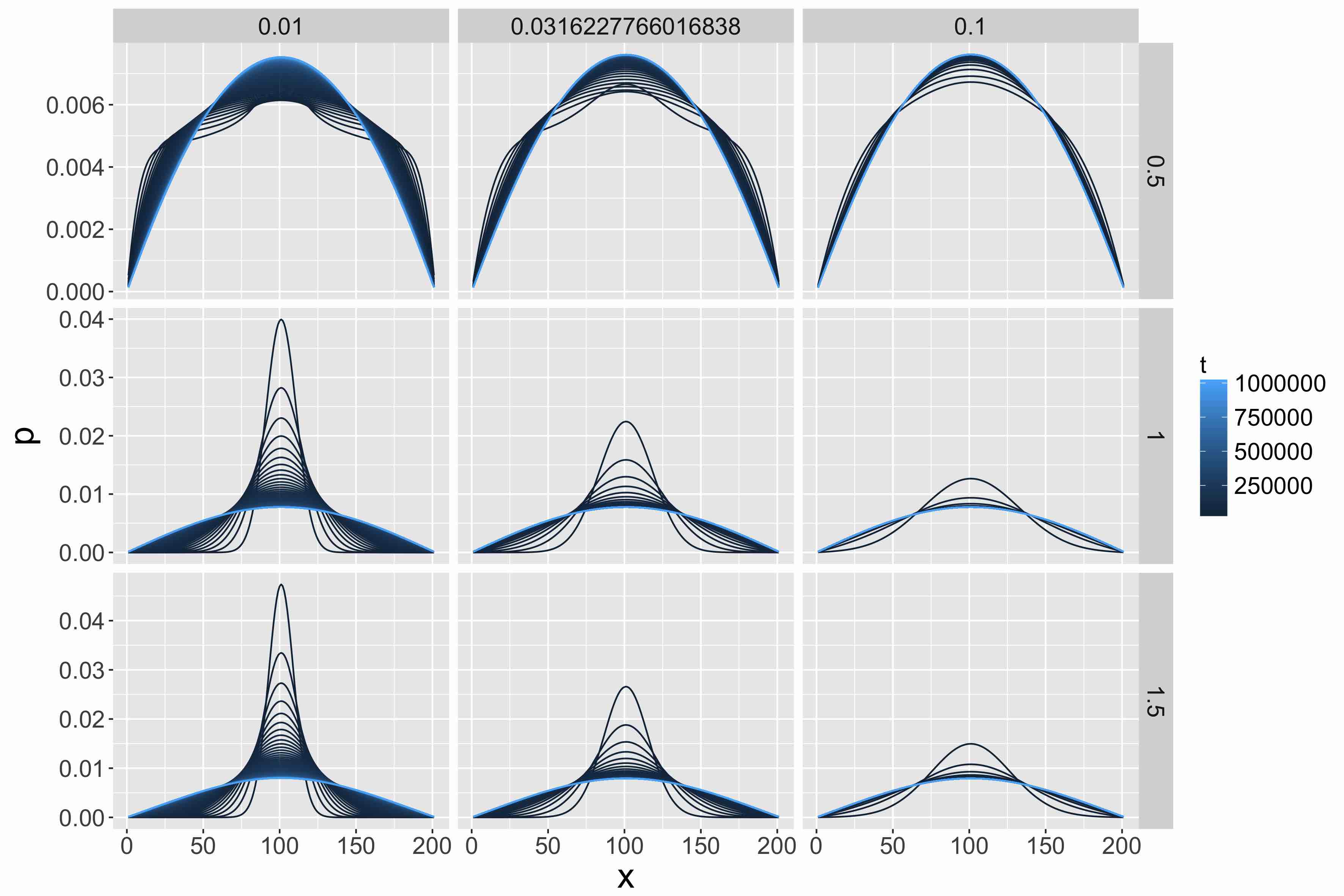}
\caption{Trajectories of densities as a function of the spatial dimension, for varying $\beta$ (columns) and $\alpha$ (rows). Color gives time.}
\label{fig:stationary}
\end{figure}

\begin{figure}[!h]
\centering
\includegraphics[width=0.49\textwidth]{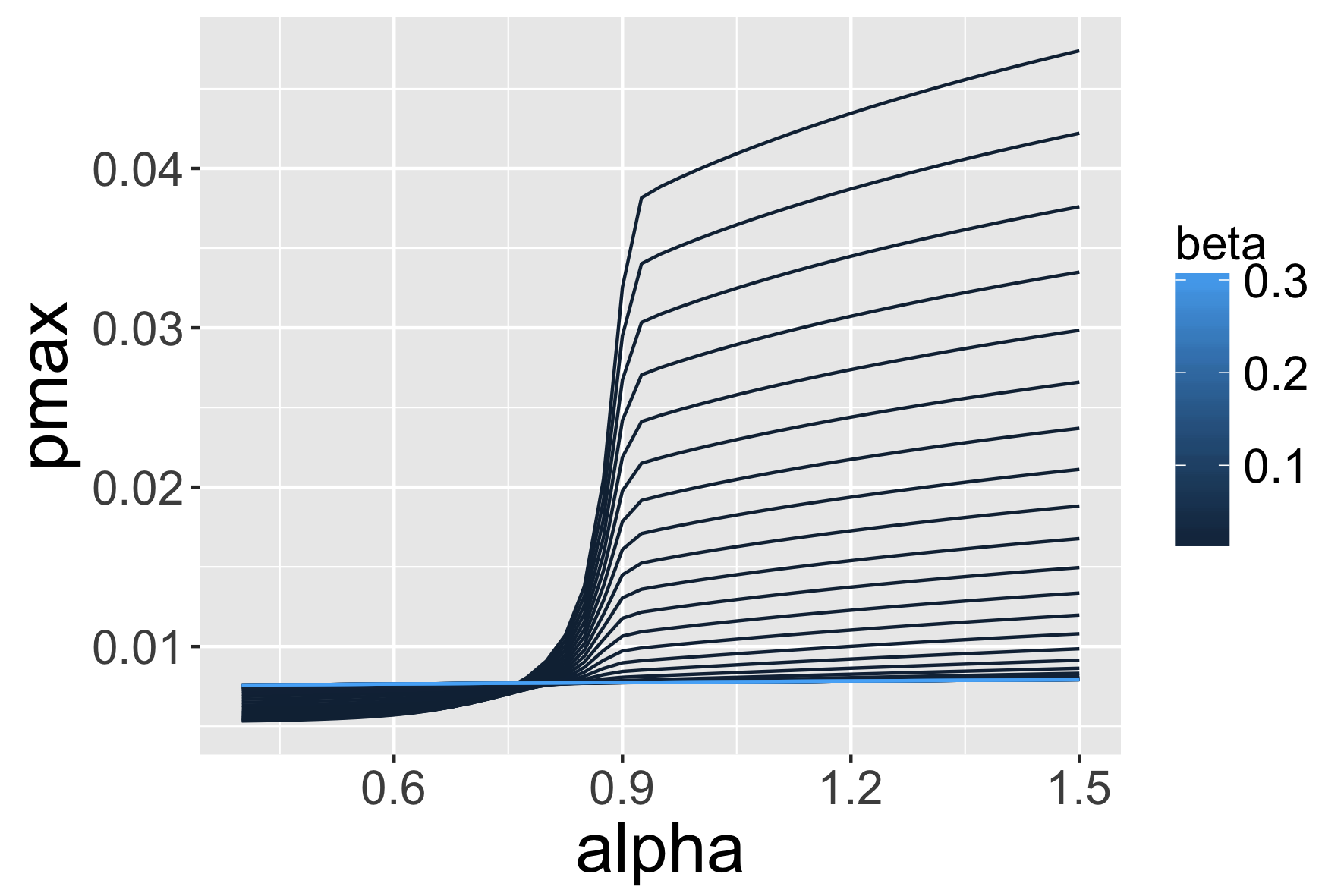}
\includegraphics[width=0.49\textwidth]{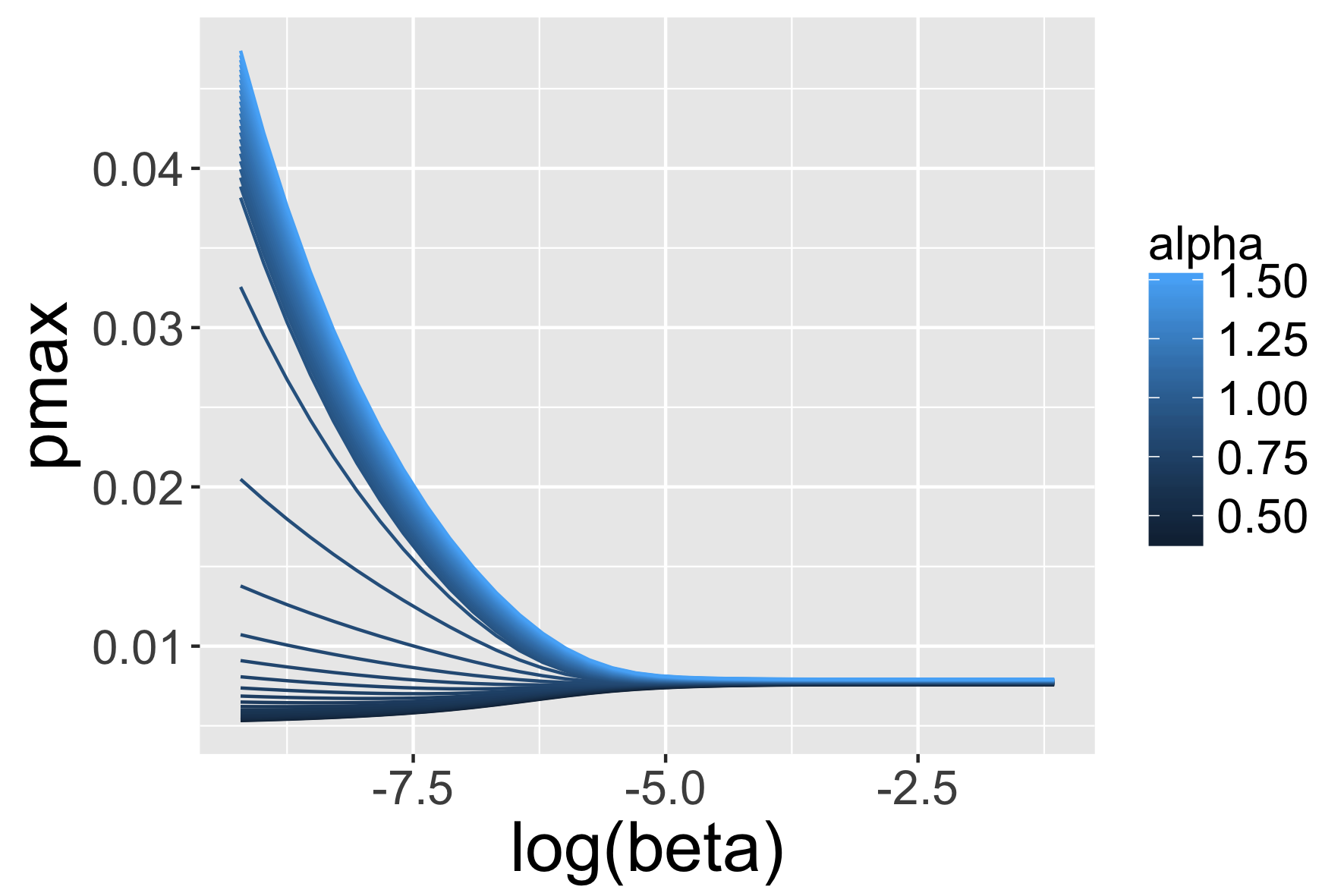}
\caption{Dependency of $\max d(t\rightarrow \infty )$ to $\alpha$ and $\beta$.}
\label{fig:pmax}
\end{figure}

The non-linearity and the integral terms making the equation above out of the scope for analytical resolution, we study its behavior numerically in some cases. Taking a simple initial condition $p_0(0)=1$ and $p_0(x)=0$ for $x\neq 0$, we show that on a finite domain, density $d(x,t)$ always converge to a stationary solution for large $t$, for a large set of values of $(\alpha,\beta)$ with fixed $N_G=10$ ($\alpha\in \left[0.4,1.5\right]$ varying with step $0.025$ and $\log\beta \in \left[-1,-0.5\right]$ with step $0.1$). We show in Fig.~\ref{fig:stationary} the corresponding trajectories on a typical subset. The variation of the asymptotic distribution as a function of $\alpha$ and $\beta$ are not directly visible, as they depend on very low values of the outward flows at boundaries. We give in Fig.~\ref{fig:pmax} their behavior, by showing the value of the maximum of the distribution. Low values of $\beta$ give an inversion in the effect of $\alpha$, whereas high values of $\beta$ give comparable values for all $\alpha$.

%%%%%%%%%%%%%%%%%%%%
%%%%%%%%%%%%%%%%%%%%

%%%%%%%%%%%%%%%%%%%%
%%%%%%%%%%%%%%%%%%%%

\end{document}